\documentclass[aps,twocolumn,prd,reprint,nofootinbib,floatfix]{revtex4-2}
\RequirePackage[l2tabu, orthodox]{nag}

\usepackage[utf8x]{inputenc}
\usepackage{microtype} 

\usepackage{import}
\usepackage{amsmath}
\usepackage{amstext, amssymb, amsthm, amsfonts, slashed}
\usepackage{physics}
\usepackage{mathtools}
\usepackage{mhchem}
\usepackage{graphicx}
\usepackage{booktabs}
\usepackage{dcolumn}
\usepackage{bm}
\usepackage{dsfont}
\usepackage[usenames,dvipsnames]{xcolor}
\usepackage[normalem]{ulem}
\usepackage{footmisc}

\usepackage{url}
\usepackage[colorlinks=true,urlcolor=blue,linkcolor=blue,citecolor=blue,hypertexnames]{hyperref}
\usepackage[nameinlink]{cleveref}
\usepackage{braket}
\usepackage{simplewick}
\usepackage{float}
\usepackage{subfigure}
\usepackage{multirow}

\usepackage{tikz}

\hypersetup{
	colorlinks,
	linkcolor={blue!75!black},
	citecolor={blue!75!black},
	urlcolor={blue!75!black}
}

\crefname{section}{Sec.\!}{Secs.\!}
\crefname{figure}{Fig.\!}{Figs.\!}
\crefname{equation}{}{}
\crefname{table}{Tab.\!}{Tabs.\!}
\crefname{appendix}{App.\!}{Apps.\!}

\usepackage{xcolor,colortbl}

\newcommand{\orcid}[1]{\href{https://orcid.org/#1}{\includegraphics[height=1.7ex,width=1.7ex]{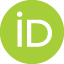}}}

\begin{document}

\title{Gravitational Waves from Confinement in $SU(N)$ Yang-Mills Theory}

\author{Stephan Huber~\orcid{0000-0002-8629-7735}}
\thanks{\href{mailto:s.huber@sussex.ac.uk}{s.huber@sussex.ac.uk}}
\affiliation{Department  of  Physics  and  Astronomy,  University  of  Sussex,  Brighton,  BN1  9QH,  U.K.}
\author{Rory~Phipps~\orcid{0009-0007-3458-4859}}
\thanks{\href{mailto:r.phipps@sussex.ac.uk}{r.phipps@sussex.ac.uk}}
\affiliation{Department  of  Physics  and  Astronomy,  University  of  Sussex,  Brighton,  BN1  9QH,  U.K.}
\author{Manuel~Reichert~\orcid{0000-0003-0736-5726}}
\thanks{\href{mailto:m.reichert@sussex.ac.uk}{m.reichert@sussex.ac.uk}} 
\affiliation{Department  of  Physics  and  Astronomy,  University  of  Sussex,  Brighton,  BN1  9QH,  U.K.}

\begin{abstract}
We provide a detailed analysis of the gravitational wave spectrum of $SU(N)$ pure Yang-Mills theory. The confinement phase transition is described with an effective Polyakov loop model, using the latest lattice data as an input. In particular, recent lattice studies clarified the large-$N$ scaling of the surface tension, which we incorporate through a modification of the kinetic term. We demonstrate that the thin-wall approximation agrees with the Polyakov loop model at small $N$ while it breaks down at large $N$. Furthermore, we include reliable estimates of the bubble wall velocity using a recently developed framework based on a large enthalpy jump at the phase transition. Altogether, this allows us to derive the gravitational wave signals for all $SU(N)$ confinement phase transitions and clarifies the behaviour at large $N$. The strongest signal arises for $N=20$, but overall the predicted signals remain rather weak. Our work paves the way for future studies of other gauge groups and systems with fermions.
\end{abstract}

\maketitle

\section{Introduction}
\label{sec:intro}
QCD-like dark sectors are interesting and well-motivated extensions of the Standard Model. Being asymptotically free at high energies, they are naturally UV-complete and, in the IR, predict fascinating bound-state dark matter candidates such as dark baryons, dark mesons, or even dark glueballs \cite{Hietanen:2013fya, Bai:2013xga, Hochberg:2014dra, Kribs:2016cew, Asadi:2021pwo, Asadi:2021yml, Asadi:2022vkc, Carenza:2022pjd, Carenza:2023eua, Bennett:2020hqd, Bruno:2024dha}. While the QCD transition is a crossover, various strongly-coupled theories exhibit a first-order confinement phase transition. The stochastic background of gravitational waves (GWs) generated by such a phase transition can be used to constrain the corresponding dark sector. It is therefore of great importance to be able to accurately compute this signal in preparation for planned and future experiments.

The study of first-order phase transitions (FOPTs) in a strongly-interacting theory is, in general, a challenging task. Due to the strong dynamics, one cannot rely on perturbation theory and instead must turn to non-perturbative methods. Non-perturbative methods can provide access to quantities of interest for the computation of the GW power spectrum. These are typically equilibrium quantities, such as the latent heat, interface tension, and pressure computed on the lattice \cite{Kogut:1983mn, Svetitsky:1983bq, deForcrand:2004jt, Lucini2005, Panero2009, Rindlisbacher:2025dqw, Bennett:2024bhy, Bennett:2025whm}, or an effective potential from functional methods \cite{Braun:2007bx, Braun:2010cy, Herbst:2010rf, Reinhardt:2012qe, Haas:2013qwp, Fister:2013bh} or holography \cite{Creminelli:2001th, Nardini2007, Gursoy:2008za, Guersoy2009, vonHarling:2017yew, Dillon2018, Bigazzi:2020phm, Agashe2020, Ares:2021ntv}. Non-equilibrium quantities necessary for the computation of the GW spectrum, such as the bubble nucleation rate and the wall velocity, are less easily accessible using these techniques. Often, such non-perturbative methods will therefore be used in tandem with an effective theory for the phase transition to compute the GW signal \cite{Baratella:2018pxi, Bigazzi:2020avc, Ares:2020lbt, Ares:2021nap, Morgante2023, Mishra:2023kiu, Helmboldt:2019pan, Huang2021, Halverson2021, Kang:2021epo, Reichert:2021cvs, Pasechnik2024, Houtz:2025ogg}.

In this work, we study pure Yang-Mills dark sectors. These are the simplest choice of a strongly coupled dark sector, and allow us to make use of the abundance of non-perturbative results, in particular lattice data. Furthermore, the order parameter for the confinement phase transition is well understood. These sectors provide dark glueballs as a dark matter candidate; however, in practice, most of the parameter space is ruled out if one insists on single-component dark matter \cite{Asadi:2022vkc, Carenza:2022pjd, Carenza:2023eua}. The only viable regime is that where the dark sector is much colder than the visible sector, but in this scenario it would constitute a very small fraction of the energy density of the Universe, resulting in a weak GW signal. Despite not being a realistic dark matter candidate, our work aims to provide a detailed understanding of the phase transition of a QCD-like hidden sector, which can be extended to more realistic dark sectors in the future. We analyse the confinement phase transition of a pure $SU(N)$ theory. This phase transition is of first order for $N \geq 3$, and the expectation value of the Polyakov loop is the order parameter. As such, the effective Polyakov loop model (PLM) \cite{Pisarski2000, Pisarski2006} is well suited to describe the thermodynamics of the phase transition. Here, we build on previous works \cite{Huang2021} where an effective tunnelling action was constructed, and the coefficients of the effective Polyakov loop potential were fitted to lattice data \cite{Panero2009, Lucini2005}.

Recent lattice results \cite{Rindlisbacher:2025dqw} demonstrated that the interface tension of the phase transition scales as $\sigma \propto N^2$ at large $N$, which is incompatible with the PLM formulated in \cite{Huang2021}. To reconcile this, we extend the setup by modifying the kinetic term in the effective Polyakov loop action, such that the model matches the results of the interface tension exactly. This leads us to a more accurate reconstruction of the Polyakov loop action, suitable for the study of bubble nucleation in the FOPT.

To determine the GW power spectrum, we use the results of numerical simulations \cite{Hindmarsh:2013xza, Hindmarsh2015, Hindmarsh:2017gnf, Giblin:2014qia, Caprini:2015zlo, Caprini:2019egz}, which express the spectrum in terms of a set of parameters, namely the nucleation temperature, duration, strength parameter, and wall velocity. While the former three quantities can be computed directly from the effective Polyakov loop action, the latter requires non-equilibrium information. In this work, we employ novel developments in the computation of the bubble wall velocity \cite{SanchezGaritaonandia2024} valid when there is a large change in the number of degrees of freedom at the transition. The error on these parameters is estimated directly from the uncertainty in the lattice results used in the fitting procedure.

The results from the effective Polyakov loop model are compared to those from the thin-wall approximation, which relies only on the interface tension and latent heat computed on the lattice. We demonstrate at small $N$ that the thin-wall approximation is well suited to determine the nucleation temperature and the duration of the phase transition. This agreement is stronger than in previous computations since our PLM matches the surface tension exactly due to the generalised kinetic term. At large $N$, the thin-wall approximation predicts a large degree of supercooling, which is a prediction outside of its domain of validity. We indeed confirm the breakdown of this approximation at large $N$, as the Polyakov loop model predicts a maximum degree of supercooling. This is due to the disappearance of the barrier in the effective potential that separates the true and false vacuum, and occurs close to the critical temperature.

With the PLM, we are able to determine the GW power spectrum at large $N$. We show that the peak amplitude has a maximum for $N = 20$ and rapidly decays as $h^2 \Omega_{\text{GW}}^{\text{peak}} \propto N^{-14/3}$ at large $N$. Overall, the signals are weak and, unfortunately, undetectable at planned and future GW observatories. This is a result of the strong coupling of the dark sector, which leads to a rapidly changing effective potential below the critical temperature. As a consequence, these transitions have a very short duration, which suppresses the signal. Given the consistency of our framework with lattice results, it can now be extended to study Yang-Mills theories of different gauge groups, or perhaps theories containing fermions in different representations, which provide more compelling dark matter candidates.

This paper is structured as follows: in \cref{sec:thin-wall} we introduce the thin-wall approximation as the simplest method with which one can estimate the GW signal. It relies only on input from the latent heat and interface tension, both of which have been computed on the lattice, but we demonstrate that its validity is restricted to small gauge groups. We improve upon the thin-wall approximation with the PLM in \cref{sec:PLM}. Here, we extend previous computations by modifying the model such that the PLM can reproduce new results of the interface tension at large $N$. In \cref{sec:GW-spectrum params} we present the results of our computations of the GW parameters, including error estimates and their large-$N$ scaling. These enter into the final computation of the GW power spectrum, the results of which we display in \cref{sec:GW-spectrum} alongside the signal-to-noise ratios at planned and future detectors. We summarise and conclude in \cref{sec:conclusions}.

\section{Thin-Wall Approximation}
\label{sec:thin-wall}
The thin-wall approximation is a common approach to obtain the relevant thermodynamic quantities for the GW spectrum from first-order cosmological phase transitions. The appeal of this approach lies in its mathematical simplicity; under the assumption of thin-walled bubbles, an analytic expression for the Euclidean action can be derived \cite{Coleman1977}. In our case, this greatly simplifies the analysis of the $SU(N)$ confinement transition, where relevant quantities such as the interface tension, $\sigma$, and latent heat, $L$, have been computed on the lattice \cite{Lucini2005, Rindlisbacher:2025dqw}.

The thin-wall approximation is an expansion in $(T-T_{\text{c}})/T_{\text{c}}$, where $T_{\text{c}}$ is the critical temperature, at which the two phases are degenerate in free energy. Naturally, the expectation is that the thin-wall approximation works well when the nucleation temperature, $T_{\text{n}}$, is close to $T_{\text{c}}$. However, whether or not the thin-wall approximation truly holds for the system of interest must be verified by a computation from the full effective potential; in our case, the effective potential of the Polyakov loop presented in \cref{sec:PLM}. Indeed, a notable result of this work is that the thin-wall approximation is insufficient for extracting the thermodynamics of the first-order confinement transition in the large-$N$ limit. Below, we outline the logic of the approximation, see also \cite{Linde:1981zj, Fuller:1987ue, Huang2021, Reichert:2021cvs, Mishra:2023kiu, Agrawal:2025wvf, Agrawal:2025xul}, and in \cref{sec:GW-spectrum A} we demonstrate its breakdown. 

The critical bubble is a non-trivial field configuration that interpolates between the true and false vacua of the system, satisfying the equation of motion and boundary conditions, as we will detail later in \cref{sec:PLM B}. If a bubble is assumed to have a thin wall, such that the bubble radius is much larger than the wall thickness, the three-dimensional Euclidean action can be approximated by
\begin{equation} \label{S_3 thin-wall def}
	S_3(T, R) = 4 \pi \sigma R^2 - \frac{4 \pi}{3}R^3 \Delta p\,,
\end{equation}
where $\Delta p$ denotes the pressure difference between the confined and deconfined phases, and $R$ is the bubble radius. In this regime, the action is assumed to be dominated by contributions from an infinitesimal bubble wall and the bulk volume, corresponding to the first and second terms of \cref{S_3 thin-wall def}, respectively. The critical radius is given by
\begin{equation} \label{Critical radius}
	R_{\text{c}} = \frac{2\sigma}{\Delta p}\,,
\end{equation}
which is obtained as the stationary point of the bubble action with respect to $R$. Substitution of this expression into \cref{S_3 thin-wall def} yields the thin-wall approximation for the energy of the critical bubble,
\begin{equation} \label{S_3 thin-wall pressure}
	S_3(T) = \frac{16 \pi}{3} \frac{\sigma^3}{\Delta p^2}\,.
\end{equation}
Expanding the pressure difference, $\Delta p \equiv \Delta p(T)$, around $T_{\text{c}}$ we find
\begin{align} \label{Pressure as latent heat}
	\Delta p (T) &\simeq \Delta p (T_{\text{c}}) + \left( T - T_{\text{c}} \right) \left. \frac{\mathrm d \left( \Delta p \right)}{\mathrm d T} \right\vert_{T=T_{\text{c}}}
	\notag \\
	&= \left( T - T_{\text{c}} \right) \left. \frac{\mathrm d \left( \Delta p \right)}{\mathrm dT} \right\vert_{T=T_{\text{c}}} \equiv L \frac{\left( T - T_{\text{c}} \right)}{T_{\text{c}}}\,,
\end{align}
where we have defined the latent heat of the transition, $L$.  Truncation of the expansion up to first-order is valid if $T \simeq T_{\text{c}}$. The Euclidean action in the thin-wall approximation then reduces to 
\begin{equation} \label{S_3 thin-wall final}
	S_3(T) = \frac{16 \pi}{3} \frac{\sigma^3}{L^2} \frac{T_{\text{c}}^2}{\left( T - T_{\text{c}} \right)^2}\,.
\end{equation}
The result in \cref{S_3 thin-wall final} is valid for any theory. For $SU(N)$ Yang-Mills theory, \cref{S_3 thin-wall final} is particularly useful since it is written entirely in terms of quantities that have been computed on the lattice for small $N$ and have associated large-$N$ fits \cite{Lucini2005, Rindlisbacher:2025dqw}. Interestingly, the lattice work \cite{Rindlisbacher:2025dqw} has demonstrated that the interface tension scales as $N^2$ in the large $N$ limit, clarifying the uncertainty of the large-$N$ scaling observed in \cite{Lucini2005}. When investigating the thin-wall approximation at large $N$ we employ the large-$N$ formulae of the interface tension and latent heat from \cite{Rindlisbacher:2025dqw},
\begin{align} \label{Large-N fits new}
	\frac{\sigma}{T_{\text{c}}^3} &= 0.0182(7)N^2 - 0.194(15)\,,
	& N &\geq 4\,, \notag\\
	\frac{L}{T_{\text{c}}^4} &= 0.360(6)N^2 - 1.88(17)\,, & N &\geq 5\,.
\end{align}
Substitution of these expressions into \cref{S_3 thin-wall final} yields an expression for the action in the thin-wall approximation that scales as $N^2$ in the large-$N$ limit,
\begin{equation} \label{S_3 thin-wall large-N 2}
	S_3^{(N \geq 5)}(T) = \frac{16 \pi}{3} \frac{\left( 0.0182N^2 - 0.194 \right)^3}{\left( 0.360N^2 - 1.88 \right)^2} \frac{T_{\text{c}}^3}{\left( T - T_{\text{c}} \right)^2}\,. 
\end{equation}
This is the expected scaling for a system that is dominated by $N^2-1$ degrees of freedom. It is important to note that the temperature-dependence of the action stems entirely from the $1/(T - T_{\text{c}})^2$ term, which is the source of the thin-wall approximation's prediction for the degree of supercooling.

\section{Polyakov Loop Model}
\label{sec:PLM}

Following the work of 't Hooft, Svetitsky, and Yaffe \cite{Hooft1978, Hooft1979, Svetitsky1982}, it has been established that the vacuum expectation value (VEV) of the Polyakov loop is an order parameter for the confinement phase transition in Yang-Mills theory at finite temperature. Pisarski subsequently pioneered the practical mean field theory approach to study confinement that forms the foundation of this work \cite{Pisarski2000, Pisarski2006}.

The first part of this section briefly motivates the Polyakov loop as an order parameter for confinement through an examination of the relevant symmetries. We then discuss the effective theory of the Polyakov loop that was used in this study, highlighting the importance of results from lattice computations in refining the model. Given the recent developments made in calculating the interface tension on the lattice \cite{Rindlisbacher:2025dqw}, this study builds on the work of \cite{Huang2021} by modifying their Polyakov loop model to match these new results. The PLM predictions for the interface tension and latent heat are then displayed with a direct comparison to the large-$N$ fits in \cite{Lucini2005} and \cite{Rindlisbacher:2025dqw}. The exact agreement observed between the lattice results and the PLM in both of these two quantities, for the entire range of $N$, demonstrates the model's ability to compute the thermodynamics of the FOPT.

\subsection{The Polyakov Loop}
In Euclidean time, $\tau$ (not to be confused with $\tau = T/T_{\text{c}}$ in later sections), at finite temperature, the temporal direction of spacetime is compactified with period $\beta = 1/T$. The gauge fields are required to obey the  periodic boundary conditions
\begin{equation} \label{Periodic b.c.}
	A_{\mu}(\tau + \beta, \textbf{x}) = A_{\mu}(\tau, \textbf{x})\,. 
\end{equation}
Under local symmetry transformations, in our case belonging to the group $SU(N)$, the gauge fields must continue to satisfy \cref{Periodic b.c.}. Given this assertion, it is easy to show that the allowed gauge transformations, applied using matrices $\Omega(\textbf{x}, \tau) \in SU(N)$, must satisfy the less stringent condition that they are periodic in Euclidean time up to an element of the centre of the gauge group:
\begin{align} \label{Twisted b.c.}
	\Omega(\tau + \beta, \textbf{x}) &= z \,\Omega(\tau, \textbf{x})\,, &  z &\in Z_N\,,
\end{align}
where $Z_N$ is the centre of $SU(N)$. Specifically, the subset of gauge transformations that are not continuously connected to the identity, due to the nontrivial topology of the temporal direction, consists of transformations that are periodic up to an element of the centre. These form a group of global transformations, known as centre transformations, that are a symmetry of the action.

\begin{table*}[t]
	\renewcommand{\arraystretch}{1.5}
	\begin{tabular}{|c|c|c|c|c|c|c|c|c|c|c|}
		\hline
		$N$ & $a_0$ & $a_1$ & $a_2$ & $a_3$ & $a_4$ & $b_3$ & $b_4$ & $b_5$ & $b_6$ & $b_8$ \\
		\hline
		3 & 6.24 & $-5.78$ & 8.67 & $-7.99$ & $-1.93$ & $- 2.26$ & 3.26 & - & - & - \\
		4 & 10.1 & $-8.85$ & 10.0 & $-12.3$ & 0.41 & - & $-1.84$ & - & 2.91 & - \\
		5 & 15.2 & $-11.6$ & 3.77 & 0.21 & $-9.68$ & - & $-9.04$ & $10.3$ & - & - \\
		6 & 22.8 & $-29.4$ & 66.9 & $-96.3$ & 29.8 & - & $-19.9$ & - & 36.9 & $-14.9$ \\
		\hline
	\end{tabular}
	\caption{Parameter values obtained from $\chi^2$-fitting of the Polyakov loop effective potential for $N = 3, 4, 5,$ and 6, using the procedure described in \cref{sec:PLM B}.}
	\label{fig: Param table}
\end{table*}

Operators that wrap around the thermal circle are sensitive to these global transformations. An example of such an operator is the Polyakov loop,
\begin{equation} \label{Polyakov loop}
	\ell(\textbf{x}) = \frac{1}{N} \text{Tr} \left[ \mathbf{L}(\textbf{x}) \right],
\end{equation}
which is defined in terms of the thermal Wilson line,
\begin{equation} \label{Thermal Wilson line}
	\mathbf{L}(\textbf{x}) = \mathcal{P} \text{exp} \left[ i g \int_0^{\beta} \! \mathrm d \tau A_0(\tau, \textbf{x}) \right],
\end{equation}
where $\mathcal{P}$ denotes path ordering, $g$ is the gauge coupling, and $A_0(\tau, \textbf{x})$ is the zeroth component of the gauge field. As the gauge field is matrix-valued, $A_{\mu}(\tau, \textbf{x}) = A_{\mu}^a(\tau, \textbf{x}) T^a$ with $T^a$ the generators of $SU(N)$ and $a = 1, 2, \dots N^2-1$, the thermal Wilson line transforms nontrivially under local symmetry transformations. Indeed, it is itself an $SU(N)$ matrix and can be shown to transform in the following way:
\begin{align} \label{Wilson line transform}
	\mathbf{L}(\textbf{x}) &\rightarrow \Omega^{\dagger}(\beta, \textbf{x}) \mathbf{L}(\textbf{x}) \Omega (0, \textbf{x}) \notag\\[1ex]
	&= z \, \Omega^{\dagger}(0, \textbf{x}) \mathbf{L}(\textbf{x}) \Omega (0, \textbf{x})\,, \ \quad \ z \in Z_N\,.
\end{align}
As the traced thermal Wilson line, the Polyakov loop is gauge invariant and can be thought of as a complex scalar field that is charged under centre transformations. From \cref{Wilson line transform} it transforms as 
\begin{align} \label{Polyakov loop transform}
	\ell(\textbf{x}) &\rightarrow z \, \ell(\textbf{x})\,, 
	&
	z &\in Z_N \,.
\end{align}
We emphasise that local gauge transformations and global centre transformations act independently. Indeed, in \cref{Wilson line transform} one can set $z = 1$ such that the transformation is purely local, and by the cyclic property of the trace, the Polyakov loop is clearly invariant under these local symmetry transformations. Conversely, a centre transformation in $SU(N)$ maps the system to one of its $N$ degenerate ground states in the deconfined phase. Although these vacua are physically equivalent, the Polyakov loop is charged under these global transformations, and it measures the extent to which this symmetry is spontaneously broken/restored.

Through its definition in terms of the thermal Wilson line, the VEV of the Polyakov loop is interpreted to be related to the free energy of a static quark, $E_{\text{q}}$, via
\begin{equation} \label{Polyakov loop VEV}
	\langle \ell \rangle = \exp( -\frac{E_{\text{q}}}{T}).
\end{equation}
Since a static quark is a colour source, in the low-temperature confined phase it takes an infinite amount of energy to isolate, such that $\langle \ell \rangle = 0$. Conversely, in the high-temperature deconfined phase, quarks are allowed states with finite free energy, such that $\langle \ell \rangle \neq 0$. Considering a Polyakov loop effective theory, the overall picture is the following: at high temperatures, the Polyakov loop has a non-zero VEV and the global centre symmetry is spontaneously broken. Once the temperature drops below the critical temperature, the Polyakov loop VEV vanishes, and the centre symmetry is spontaneously restored. Thus, the VEV of the Polyakov loop acts as an order parameter for the confinement phase transition in pure Yang-Mills theory. Such an effective theory can, in principle, be derived directly from the action of the full theory. As discussed in \cref{sec:PLM B}, an alternative approach is to write down a simple effective potential in terms of free parameters which preserves the $Z_N$ symmetry, and fit the potential to lattice data \cite{Helmboldt:2019pan, Huang2021, Halverson2021, Kang:2021epo, Reichert:2021cvs, Pasechnik2024, Houtz:2025ogg, Ratti2006}.

It should be noted that the introduction of dynamical quarks in the fundamental representation explicitly breaks centre symmetry, such that the Polyakov loop is no longer an exact order parameter. More precisely, a general centre transformation on fundamental quarks violates the antiperiodic boundary condition they must satisfy at finite temperature. In this scenario, the transition remains of first order only if the quark masses are significantly larger than the confinement scale \cite{Saito2011, Aarts2023, Ratti2006}. This is important in QCD-like theories. In theories with dynamical quarks in certain higher representations, the Polyakov loop remains a strict order parameter \cite{Sannino2005}.

\subsection{Polyakov Loop Effective Action}
\label{sec:PLM B}

The effective action written in terms of the traced thermal Wilson line can be directly derived from the Yang-Mills action of the theory. Explicit analytic computations inevitably require approximations and perturabtive expansions. After the work of Svetitsky and Yaffe \cite{Svetitsky1982}, many attempts were made \cite{Polonyi1982, Gross1984, Ogilvie1984, Matsuoka1984, Green1984}. See \cite{Fukushima2017} for a review on this subject.

The physics of the confinement transition is inherently non-perturbative. Therefore, perturbative approaches will likely not suffice and input from lattice results is required. Here, we use an effective potential which preserves the $Z_N$ symmetry in terms of a set of free parameters, and fix these parameters using results obtained from lattice computations, see also \cite{Huang2021} upon which our work is based. The simplest potential one can write down that preserves the $Z_N$ symmetry and exhibits a FOPT reads
\begin{equation} \label{Potential general ansatz}
	V_{\text{eff}}^N(\ell, T) = T^4 \left( A \left| \ell \right|^2 + B \left| \ell \right|^4 + C \left( \ell^N + \ell^{*N} \right)+ \dots \right), 
\end{equation}
where the factor of $T^4$ ensures the correct dimensionality of the potential. Note that we are following a strict expansion in powers of $\ell$ and the term $(\ell^N + \ell^{*N})$ is only included if $N$ is smaller or equal to our overall expansion order. While the dimensionless coefficients $A, B,$ and $C$ are generally temperature-dependent, in \cite{Ratti2006, Fukushima2017} it is reasoned that only the coefficient $A$ needs to be temperature-dependent to capture the necessary effects. We explicitly verified this assumption by finding a negligible improvement in the fit with temperature-dependent higher-order coefficients. For instance, in the $SU(4)$ effective potential we promoted $b_4 \rightarrow a_5 + a_6(T_{\text{c}}/T)$ and re-fitted. We found only a marginal improvement in the $\chi^2$ value of the fit and a qualitatively identical potential, in particular in relation to its behaviour with temperature.

\begin{figure}[t]
	\includegraphics[width=\linewidth]{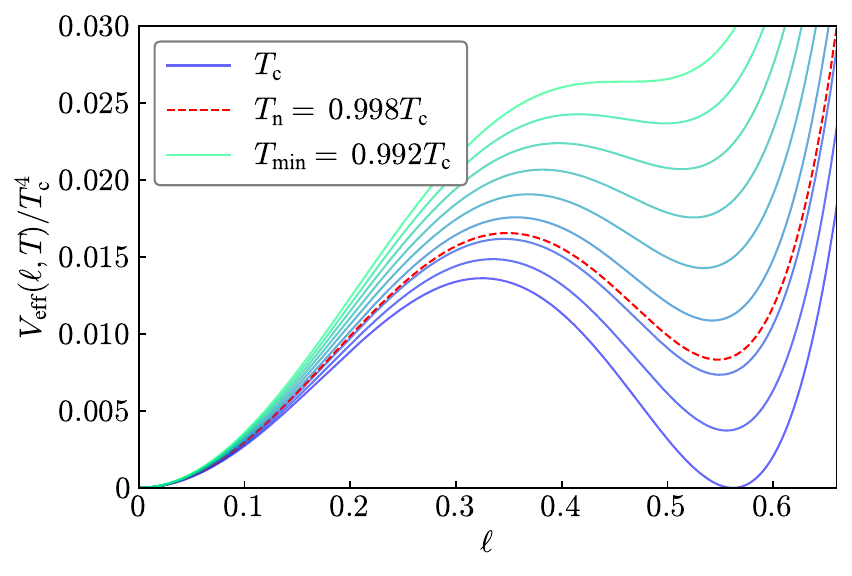}
	\caption{Reconstructed $SU(4)$ Polyakov loop effective potential, in units of $T_\text{c}^4$, plotted as a function of the Polyakov loop along the real direction. Each curve was obtained for a given temperature, ranging from the critical temperature, $T_\text{c}$, to $T_{\text{min}}$; the temperature at which the barrier disappears.}
	\label{fig: SU4 potential}
\end{figure}

The VEV of the Polyakov loop is, in general, a complex number; however, the $Z_N$ symmetry of the potential allows one to choose a convenient direction. Denoting by $\ell_0$ the magnitude of the VEV of the Polyakov loop, which is temperature-dependent, it is easy to see that
\begin{align} \label{Z_N vacua}
	\langle \ell \rangle &= \ell_0 \exp( i \frac{2 \pi j}{N}), 
	&
	j &= 0, 1, \dots, N-1\,,
\end{align} 
for temperatures above the critical. Without loss of generality, we can choose $j = 0$, which simplifies the form of the potential in \cref{Potential general ansatz} as $\ell$ is strictly real, such that $ \ell = \ell^*$. For the deconfinement-confinement transition rate, it is sufficient to study tunnelling from one representative of the deconfined vacua since all $Z_N$ deconfined vacua are degenerate. Therefore, the following ansatz for the Polyakov loop effective potential was used:
\begin{equation} \label{Potential temperature ansatz}
	V_{\text{eff}}^N(\ell, T) = T^4 \left( -\frac{b_2(T)}{2}  \ell^2 + \sum_{i = 3}^{i_\text{max}} b_i  \ell^i \right), 
\end{equation} 
where we have rewritten the polynomial coefficients of \cref{Potential general ansatz} in terms of parameters $b_i$, and
\begin{equation} \label{Temperature dependent coeff}
	b_2(T) = a_0 + a_1 \left( \frac{T_{\text{c}}}{T} \right) + a_2 \left( \frac{T_{\text{c}}}{T} \right)^{\!2} + a_3 \left( \frac{T_{\text{c}}}{T} \right)^{\!3} + a_4 \left( \frac{T_{\text{c}}}{T} \right)^{\!4}.
\end{equation}
For our purposes, truncating the sum at $i_{\text{max}} = 8$ was sufficient. We verified that higher-order terms were negligible and so dropped them both for simplicity and to reduce exposure to overfitting errors. Note that for a given value of $N$, not all terms in the effective potential in \cref{Potential temperature ansatz} contribute. Firstly, the required $Z_N$ symmetry forces certain terms to be exactly zero; for instance, $b_3=b_5=b_7=0$ in the $SU(6)$ potential. Additionally, in certain  small-$N$ cases, higher-order terms were found to be irrelevant and so were dropped for simplicity. This is shown in \cref{fig: Param table}.

\begin{figure}[t]
	\includegraphics[width=\linewidth]{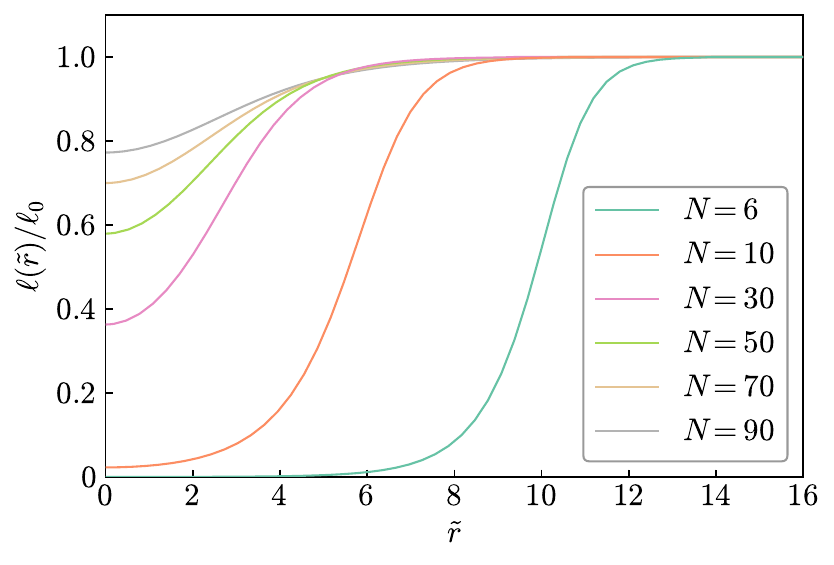}
	\caption{Bounce solution obtained at the nucleation temperature, scaled by the Polyakov loop VEV, $\ell_0 (T = T_{\text{n}})$, and plotted as a function of the dimensionless radius, $\tilde{r} = r T_\text{c}$, for various values of $N$. A critical temperature of $T_{\text{c}}=1$ GeV was used.}
	\label{fig: Large-N bounce}
\end{figure}

In \cite{Huang2021}, each free parameter was obtained by fitting the potential in \cref{Potential temperature ansatz} to lattice results of the pressure and the trace of the energy-momentum tensor as a function of temperature \cite{Panero2009} for $N = 3, 4, 5, 6,$ and $8$. The $\chi^2$-minimisation was subject to the constraint $\langle \ell \rangle \rightarrow 1$ when $T \rightarrow \infty$, as well the Stefan-Boltzmann limit, $p/T^4|_{T \rightarrow \infty} \rightarrow 1.21 \cdot (N^2 - 1) \cdot \pi^2/45$ \cite{Panero2009}.

Here, we improve the fitting procedure in two ways. Firstly, we include the latest latent heat results from the lattice in the fitting procedure. We employ the infinite-volume-extrapolated values of the latent heat from \cite{Giusti:2025fxu} and \cite{Rindlisbacher:2025dqw} for $N = 3$ and $N = 4, 5, 8$,  respectively, whereas for $N = 6$ the large-$N$ formula in \cref{Large-N fits new} was used. As an example, we show the $N=4$ Polyakov loop effective potential obtained using this fitting method in \cref{fig: SU4 potential}. To ascertain the significance of the introduction of the latent heat into the fitting procedure, one can compare the parameter values displayed in \cref{fig: Param table} to those obtained in the work of \cite{Huang2021} (noting the different conventions in defining the coefficients). The largest change in the parameter values occurred for the $SU(3)$ effective potential, which can be attributed to the significantly different value for the latent heat obtained in the recent study of \cite{Giusti:2025fxu} as compared to that predicted by the data in \cite{Panero2009}. The incompatibility of these data is a source of uncertainty discussed in \cref{sec:Gw-spectrum-errors}. The parameters of the $N = 4, 5$ and 6 effective potentials changed only slightly.

Secondly, we improve the treatment of the kinetic term in the Polyakov loop effective action. In \cite{Huang2021, Halverson2021}, a canonical kinetic term for the thermal Wilson line was assumed; however, as we will show, this results in a prediction for the interface tension which does not match the $N^2$ scaling displayed in \cref{Large-N fits new}. Remaining ignorant about the form of the kinetic term and imposing spherical symmetry, the Polyakov loop three-dimensional Euclidean action reads
\begin{equation} \label{S_3 def}
	S_3(T) = 4 \pi T \int_0^{\infty \!} \mathrm dr' r'^2 \left[ \frac{Z_{\ell}}{2} \left( \frac{\mathrm d \ell}{\mathrm d r'} \right)^{\!2} + V_{\text{eff}}'(\ell, T) \right],
\end{equation}
where $r' = r T$, with $r$ the radial coordinate, $V_{\text{eff}}'(\ell, T) = V_{\text{eff}}/T^4$, and $Z_{\ell}$ is in general some undetermined function of the Polyakov loop and the temperature, which is taken to be field- and temperature-independent here. For a canonical kinetic term, $Z_{\ell}=1$. To obtain the bubble profile, we solve the Euler-Lagrange equation of motion,
\begin{equation} \label{EL EoM}
	\frac{\mathrm d^2 \ell}{\mathrm dr'^2} + \frac{2}{r'} \frac{\mathrm d \ell}{\mathrm d r'} - \frac{1}{Z_{\ell}} \frac{\partial V_{\text{eff}}'(\ell, T)}{\partial \ell} = 0\, ,
\end{equation}
subject to the boundary conditions
\begin{align} \label{EoM b.c.s}
	\left. \frac{ \mathrm d \ell}{\mathrm  d r'} \right\vert_{r' = 0} &= 0\,, 
	&&\text{and} &
	\ell(r' \rightarrow \infty) &= \ell_{0}\,.
\end{align}
Eq.~\cref{EL EoM} is solved numerically using a shooting method, and \cref{fig: Large-N bounce} displays the obtained critical bubble profiles for increasing $N$.

The interface tension is well defined only at the critical temperature, and is given by
\begin{equation} \label{Sigma def}
	\sigma = T_{\text{c}}^3 \int_{-\infty}^{\infty} \mathrm dz' \left[ \frac{Z_{\ell}}{2} \left( \frac{\mathrm d \ell}{\mathrm dz'} \right)^{\!2} + V_{\text{eff}}'(\ell, T_{\text{c}}) \right].
\end{equation}
At the critical temperature, the second term of \cref{EL EoM} (replacing $r' \rightarrow z'$ as we are considering a planar wall) vanishes, and the resulting differential equation has a domain wall solution with $\ell(z' \rightarrow -\infty) =0$ and $\ell(z' \rightarrow \infty) =  \ell_{0}$. We can then rewrite \cref{Sigma def} as
\begin{equation} \label{Sigma sqrt}
	\sigma = T_{\text{c}}^3 \int_0^{\ell_0} \mathrm d \ell' \sqrt{2 Z_{\ell} V_{\text{eff}}'(\ell', T_{\text{c}})}\,.
\end{equation}
Most remarkably, \cref{Sigma sqrt} relates the surface tension, which is a static quantity measurable on the lattice, with the prefactor $Z_{\ell}$ of the kinetic term. The latter is a dynamic quantity and is therefore hard to access on the lattice. The relation \cref{Sigma sqrt} bridges this gap at least at the critical temperature. Determining the coefficient $Z_{\ell}$ by appropriately matching to lattice data of the interface tension is the significant improvement of this work in terms of deriving PLM parameters.

Lattice results have demonstrated that the pressure, $\Delta p = -\Delta V_{\text{eff}}$, scales as $N^2$ in the large-$N$ limit. This justifies the assumption used in this work that we can access the large-$N$ behaviour by a rescaling of the fitted Polyakov loop potential at the largest $N$ accessible with the lattice. Here, we rescale the $N=6$ potential, and we confirmed that the potential rescaled to $N=8$ accurately matches the $N=8$ lattice data. This demonstrates that $N=6,8$ are already in the large-$N$ regime.

More accurately, we rescale the potential such that the large-$N$ fit of the latent heat in \cref{Large-N fits new} is matched. To ensure the PLM reproduces the exact values of the latent heat from the large-$N$ fit for all $N > 6$, one must scale the effective potential in the following way:
\begin{equation} \label{V scaling}
	V_\text{eff}^{N>6} = \left[\frac{0.360N^2 - 1.88}{0.360 \cdot 6^2 - 1.88} \right] \cdot V_{\text{eff}}^{N=6}.
\end{equation}
If one were to scale the effective potential as $N^2/6^2$, at large $N$ the PLM would estimate the latent heat to be a factor of $ \sim 0.85$ smaller than the large-$N$ fit of \cref{Large-N fits new}.

\begin{table}[t]
	\renewcommand{\arraystretch}{1.5}
	\begin{tabular}{|c|c|c|c|}
		\hline
		$N$ & $L/T_{\text{c}}^4$ & $\sigma/{T_{c}^3}$ & $\delta(N)$ \\
		\hline
		3 & 1.175(10) & 0.0200(6) & 0.14 \\
		4 & 3.8(2) & 0.0997(76) & 0.17 \\
		5 & 7.2(2) & 0.258(11) & 0.20 \\
		6 & 11.08(27) & 0.461(29) & 0.14 \\
		8 & 21.5(4) & 0.994(47) & 0.19 \\ 
		$ \rightarrow \infty$ &0.360(6)$N^2$&0.0182(7)$N^2$  & $\rightarrow 0.24$ \\ 
		\hline
	\end{tabular}
	\caption{Tabulated latent heats from \cite{Giusti:2025fxu} ($N = 3$) and \cite{Rindlisbacher:2025dqw} ($N = 4, 5, 6,$ and $8$) and interface tensions from \cite{Lucini2005} ($N=3$) and \cite{Rindlisbacher:2025dqw} ($N=4, 5, 6,$ and $8$), with corresponding $\delta$-parameter from the Polyakov loop kinetic term obtained by matching the solution of \cref{Sigma sqrt} to the data for each $N$. While the interface tensions for $N=3, 4, 5,$ and $8$ were matched to direct lattice results, that for $N=6$ was equated to the large-$N$ fit of \cref{Large-N fits new}.}
	\label{fig: Sigma table}
\end{table}

\begin{figure*}[t]
	\includegraphics[width=0.49\textwidth]{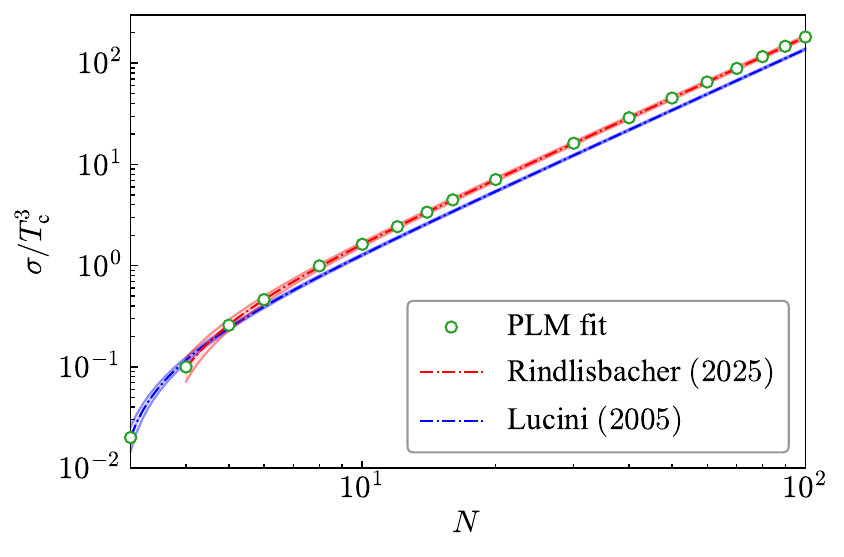}
	\hfill
	\includegraphics[width=0.4875\textwidth]{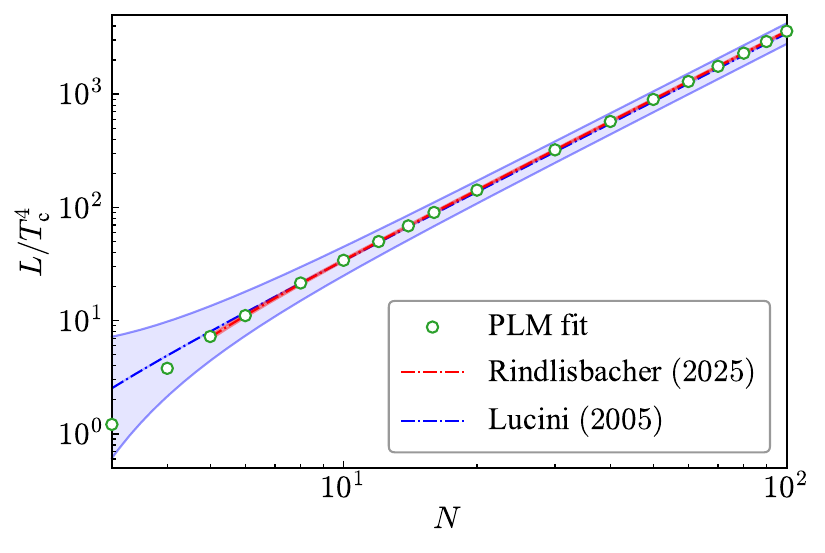}
	\caption{Interface tension in units of $T_\text{c}^3$ (left), and latent heat in units of $T_\text{c}^4$ (right), plotted as a function of the number of colours. The red and blue curves were drawn using large-$N$ fits of the interface tension and latent heat obtained from two separate lattice computations, \cite{Rindlisbacher:2025dqw} and \cite{Lucini2005}, respectively. The green points were calculated from the PLM fit using the formulae for the interface tension \cref{Sigma sqrt} and latent heat \cref{Pressure as latent heat}. Both of these quantities, by definition, were computed at the critical temperature.}
	\label{fig: Sigma and Lh plots}
\end{figure*}

Using the above fact, one can obtain the large-$N$ scaling of the interface tension from its expression in \cref{Sigma sqrt} for a given choice of $Z_{\ell}$. If one would assume that the Polyakov loop has a canonical kinetic term ($Z_\ell=1$), the interface tension would scale like $\sigma \propto N$ in the large-$N$ limit. This disagrees with the observations of \cite{Rindlisbacher:2025dqw}. To recover the desired scaling, we set $Z_\ell = \delta N^2$, where $\delta$ is some coefficient to be tuned such that the interface tension from \cref{Sigma sqrt} agrees exactly with the results from the lattice, and is taken to be temperature-independent. The results in \cite{Rindlisbacher:2025dqw} of the direct computations of the interface tension for $N=4, 5,$ and $8$ at infinite volume were used in this matching procedure, as well as that for $N=6$ from the large-$N$ fit of \cref{Large-N fits new}. In the case of $N=3$, the infinite-volume computation of the interface tension from \cite{Lucini2005} was used. For $N > 8$, $\delta(N)$ was obtained by matching the large-$N$ expression for the interface tension in \cref{Large-N fits new} to \cref{Sigma sqrt} using the rescaled $N=6$ effective potential, as is discussed above. The values of $\delta$ for each $N$ are tabulated in \cref{fig: Sigma table}. Remarkably, the parameter $\delta$ is an almost $N$-independent number of order one, which highlights that the $N^2$ factor captures most of the non-trivial scaling.

In summary, the Euclidean action used in this work then reads
\begin{equation} \label{S_3 final}
	S_3(T) = 4 \pi T \int_{0}^{\infty} \mathrm dr' r'^2 \left[ \frac{\delta N^2}{2} \left( \frac{\mathrm d \ell}{\mathrm d r'} \right)^2 + V_{\text{eff}}'(\ell, T) \right].
\end{equation}
We have justified the introduction of $Z_\ell = \delta N^2$ in front of the kinetic term from considerations of the scaling of the interface tension with $N$. It is important to note that this scaling of the action, and thus the kinetic term, with $N$ has been observed elsewhere in the literature \cite{Fukushima2017}, including in holographic studies of confinement \cite{Nardini2007, Agashe2020, Gursoy:2008za, Guersoy2009, Dillon2018, vonHarling:2017yew, Baratella:2018pxi, Bigazzi:2020phm, Ares:2021ntv, Mishra:2023kiu} where it is a general feature. Crucially, the scaling of the action with $N$ from the thin-wall approximation \cref{S_3 thin-wall large-N 2} is also in agreement.

\subsection{Interface Tension and Latent Heat}

We show that the Polyakov loop model accurately matches lattice observables, namely the interface tension and latent heat as a function of $N$. In \cref{fig: Sigma and Lh plots}, we plot the interface tension (left) and latent heat (right) against $N$, both from the large-$N$ formulae of \cite{Rindlisbacher:2025dqw} and \cite{Lucini2005}, and from the PLM. In both cases, the output of the PLM and lattice fitted formulae are in excellent agreement, particularly with the results of \cite{Rindlisbacher:2025dqw}. The fitting procedure and $\delta$-parameter tuning were employed with this goal in mind, and this acts as a check that the PLM can reproduce the thermodynamics of the phase transition. Note that the weaker agreement between the large-$N$ fit of the latent heat and the PLM for $N = 3, 4$ and $5$ is because the large-$N$ fits are not applicable at small values of $N$, and are therefore not used.

By definition, both the interface tension and latent heat are computed at the critical temperature. Away from the critical temperature, the fitting procedure utilises the pressure and the trace of the energy-momentum tensor computed on the lattice \cite{Panero2009}. Crucially, the $N = 3, 4, 5,$ and 6 reconstructed effective potentials displayed similar behaviour around $T_{\text{c}}$, which gives some confidence that the effective theory correctly models the system dynamics at temperatures below $T_{\text{c}}$, at least for small $N$. As discussed later, there is greater uncertainty at large $N$.

Given the error estimates associated with the interface tension and latent heat computed on the lattice, the matching procedure described above allows for an upper- and lower-bound on the GW power spectrum to be obtained. Specifically, we re-fitted the effective potential by instead using the upper- and lower-bounds of the latent heat as the fitting constraint, giving effective potentials which reproduce these values. Then, we tuned $\delta$ to match the upper- and lower-bounds of the interface tension, in total giving four cases. In the results for the GW parameters, as described in \cref{sec:GW-spectrum params}, we represent the maximal and minimal values obtained from these four cases as an error bar.

\begin{figure*}[t]
	\includegraphics[width=0.49\textwidth]{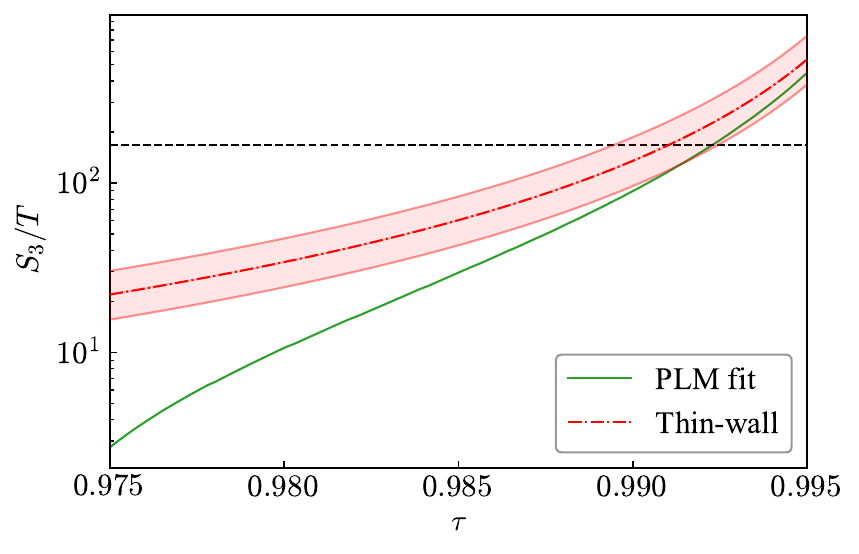}
	\hfill
	\includegraphics[width=0.49\textwidth]{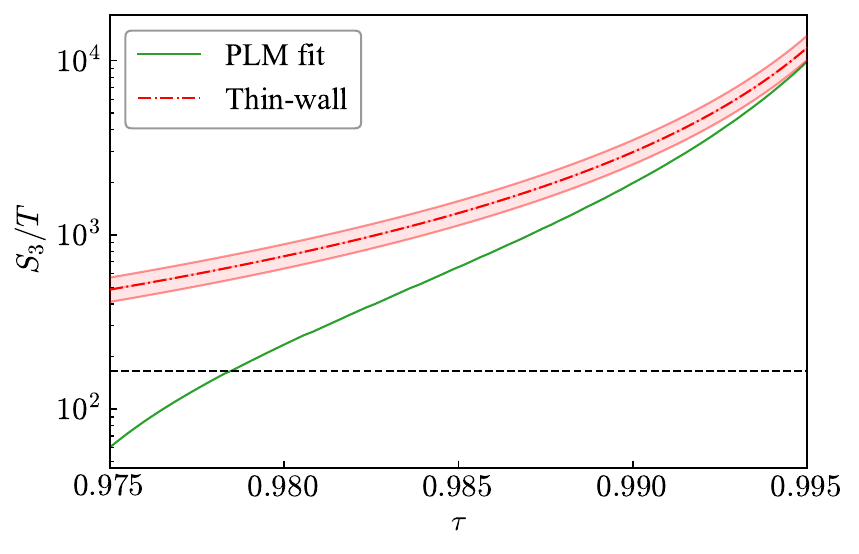}
	\caption{$S_3/T$ plotted as a function of temperature for $N=6$ (left) and $N=20$ (right). The $\tau$-axis runs from slightly above $\tau_{\text{min}}$ to slightly below $\tau = 1$. The black dashed line represents the point at which the nucleation condition is met, $\Gamma/H^4 \sim 1$. For $N=6$, the thin-wall approximation and the PLM fit show a reasonable agreement at the nucleation temperature, $\tau_n$, while for $N=20$, the thin-wall approximation is clearly breaking down.}
	\label{fig: S_3/T plots}
\end{figure*}

\section{GW Parameters}
\label{sec:GW-spectrum params}
We outline the procedures undertaken to compute the stochastic GW background generated from a first-order confinement phase transition in $SU(N)$ Yang-Mills theory, using the PLM introduced above. The GW power spectrum receives contributions from three independent sources: collisions of expanding bubbles \cite{Kosowsky:1992rz, Kosowsky:1991ua, Kosowsky:1992vn, Kamionkowski:1993fg, Caprini:2007xq, Huber:2008hg, Weir:2016tov, Jinno:2016vai}, sound waves \cite{Hindmarsh:2013xza, Hindmarsh2015, Hindmarsh:2016lnk, Hindmarsh:2017gnf, Ellis2020}, and magnetohydrodynamic (MHD) turbulence \cite{Caprini:2009yp, Kosowsky:2001xp, Dolgov:2002ra, Nicolis:2003tg, Caprini:2006jb, Gogoberidze:2007an, Kahniashvili:2008pe, Kahniashvili:2009mf, Kisslinger:2015hua, Hindmarsh2021}. In the case of thermal FOPTs such as those studied in this work, the contribution from sound waves dominates. We checked explicitly that this holds in our computation. Numerical simulations of GWs generated from acoustic sources were performed in \cite{Hindmarsh2015}, the results of which lead to the following fit:
\begin{equation} \label{GW PS def}
	h^2 \Omega_{\text{GW}}(f) = h^2 \Omega_{\text{GW}}^{\text{peak}} \left( \frac{f}{f_{\text{peak}}} \right)^{\!3} \left( \frac{7}{4 + 3(f/f_{\text{peak}})^2} \right)^{\!7/2},
\end{equation}
where $h =H/(100 \text{km/s/Mpc}) $ is the dimensionless Hubble parameter. The peak frequency of the power spectrum, $f_{\text{peak}}$,  is given by
\begin{equation} \label{GW peak freq}
	f_{\text{peak}} = 1.9 \cdot 10^{-5} \hspace{0.05cm} \text{Hz} \left( \frac{\tilde{\beta}}{\xi_{\text{w}}} \right) \left( \frac{T_{\text{n}}}{100 \hspace{0.05cm} \text{GeV}} \right) \left( \frac{g_*}{100} \right)^{\!1/6},
\end{equation}
and the peak amplitude, $h^2 \Omega_{\text{GW}}^{\text{peak}}$, by
\begin{equation} \label{GW peak amp}
	h^2 \Omega_{\text{GW}}^{\text{peak}} = 2.65 \cdot 10^{-6} \left( \frac{\xi_{\text{w}}}{\tilde{\beta}} \right) \left( \frac{\lambda_{\alpha} \kappa \alpha}{1 + \alpha} \right)^{\!2} \left( \frac{100}{g_*} \right)^{\!1/3},
\end{equation}
see \cite{Caprini:2015zlo}. The nucleation temperature, $T_{\text{n}}$, effective number of relativistic degrees of freedom, $g_*$, inverse duration, $\tilde{\beta}$, bubble wall velocity, $\xi_{\text{w}}$, strength parameter, $\alpha$, and efficiency factor, $\kappa$, are defined in the following sections. The computation of the GW power spectrum is thus reduced to a determination of each of these parameters. The term $\lambda_\alpha$ encodes the effect of the presence of visible matter on the observed power spectrum from the dark sector phase transition, and is given by
\begin{equation} \label{Omega SUN}
	\lambda_\alpha =  \frac{g_*^{SU(N)}}{g_*}= \frac{g_*^{SU(N)}}{g_*^{\text{SM}} + g_{*}^{SU(N)}}\,.
\end{equation}
In \cref{Omega SUN}, $g_*^{\text{SM}}$ denotes the Standard Model (SM) degrees of freedom, which are temperature dependent, and $g_*^{SU(N)} = 2 \cdot (N^2 - 1)$ denotes that from the $SU(N)$ dark sector. At large $N$, $g_*^{SU(N)} \gg g_*^{\text{SM}}$ and $\lambda_\alpha \simeq 1$. It is common in the literature for \cref{Omega SUN} to be absorbed in a redefinition of $\alpha$ \cite{Breitbach2019, Fairbairn2019, ArcherSmith2020}. 

To quantify the detectability of a GW signal, we employ the signal-to-noise ratio (SNR). At a detector with sensitivity curve $h^2 \Omega_{\text{det}}(f)$, and $h^2 \Omega_{\text{GW}}(f)$ defined in \cref{GW PS def}, the SNR is given by
\begin{equation} \label{SNR def}
	\text{SNR} = \sqrt{\frac{T}{s} \int_{f_{\text{min}}}^{f_{\text{max}}} \mathrm df \left( \frac{h^2 \Omega_{\text{GW}}}{h^2 \Omega_{\text{det}}} \right)^{\!2}}\,,
\end{equation}
where $T$ is the observation period of the detector in seconds. In this work, we take $T = 3$ years, and assume an SNR $>1$ is detectable.

\subsection{Nucleation Temperature, $T_{\text{\normalfont{n}}}$}
\label{sec:GW-spectrum A}
We use the nucleation temperature, $T_{\text{n}}$, to estimate the temperature of GW production. The nucleation temperature is defined to be the temperature at which the bubble nucleation rate, $\Gamma$, per Hubble volume per Hubble time is approximately equal to one \cite{Caprini:2015zlo}, i.e., $\Gamma/H^4 \sim 1$, with \cite{Linde:1981zj}
\begin{equation} \label{Nucleation rate def}
	\Gamma(T) = T^4 \left( \frac{S_{3}(T)}{2 \pi T} \right)^{\!3/2} e^{-S_3(T)/T}.
\end{equation}
The Hubble parameter is obtained from the Friedmann equation under the assumption that the transition occurs in the radiation-dominated era,
\begin{equation} \label{Hubble param}
	H(T) = \sqrt{\frac{\pi^2 g_{*}}{90}}\, \frac{T^2}{M_{\text{P}}}\,.
\end{equation}
Above, $M_{\text{P}} = 2.435 \cdot 10^{18}$ GeV is the reduced Planck mass. In general, a more accurate definition of the temperature of GW production is the percolation temperature, $T_{\text{p}}$ \cite{Guth1980, Guth1981}. However, since in our case the FOPT occurs rapidly, $T_{\text{p}} \simeq T_{\text{n}}$ holds and the nucleation temperature is a sufficient estimate.

\begin{figure}[t]
	\includegraphics[width=\linewidth]{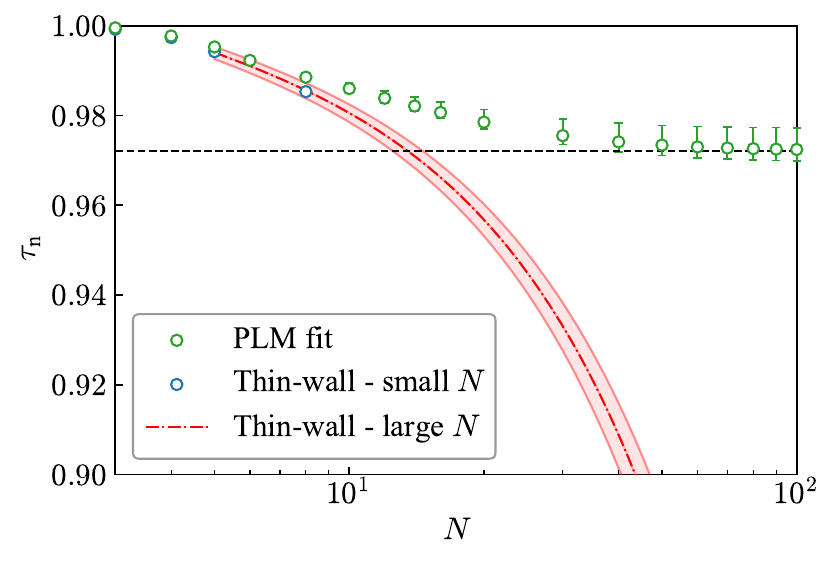}
	\caption{Nucleation temperature, in units of the critical temperature ($T_{\text{c}}=1$ GeV), plotted as a function of the number of colours. The red curve was obtained using the thin-wall approximation, see \cref{S_3 thin-wall large-N 2}. The blue points were obtained using the thin-wall formula \cref{S_3 thin-wall final} and the infinite-volume extrapolated values of the interface tension and latent heat given in \cite{Rindlisbacher:2025dqw}. The green points are the nucleation temperatures computed directly from the PLM. The black dashed line displays the temperature at which the barrier disappears for the $N \geq 6$ effective potentials, $\tau_{\text{min}}^{N=6}$.}
	\label{fig: Nucleation temp plot}
\end{figure}

From \cref{Nucleation rate def}, the quantity of primary importance in the determination of $T_{\text{n}}$ is $S_3/T$. In \cref{fig: S_3/T plots}, we plot $S_3/T$ for $N=6$ and $N=20$ against $\tau$, with
\begin{align}
	\tau = T/T_{\text{c}}\,,
\end{align}
using both the thin-wall approximation and the PLM. Near the critical temperature, these curves are in agreement, as one would expect given that the critical bubble configuration is thin for temperatures just below $T_{\text{c}}$. Since it diverges both at $T = 0$ and $T = T_{\text{c}}$, the thin-wall formula for $S_3/T$ from \cref{S_3 thin-wall final} has a minimum at some temperature in the range $0 < T < T_{\text{c}} $. Conversely, the fitted Polyakov loop effective potentials from \cref{Potential temperature ansatz} exhibit a temperature $0 < T_{\text{min}} < T_{\text{c}}$ at which $S_3/T$ vanishes. This is the temperature below which the deconfined phase no longer exists as a metastable state. As in the thin-wall approximation, $S_3/T$ from the PLM also diverges at $T = T_{\text{c}}$, and it is therefore a monotonically increasing function of $T$. This demonstrates that the curves in \cref{fig: S_3/T plots}
have to disagree at lower temperatures. Notably, the existence of a minimum temperature was also observed with other methods including holography for the confinement phase transition \cite{Gursoy:2008za, Guersoy2009, Buchel:2021yay, Morgante2023, Mishra:2024ehr, Agrawal:2025xul}.

To show that the thin-wall approximation breaks down in the large-$N$ limit, we highlight two pieces of evidence. Firstly, consider the profile of the PLM critical bubble as $N$ increases in \cref{fig: Large-N bounce}. From the scaling of the interface tension, which resulted in a kinetic term scaling with $N^2$, the three-dimensional action of \cref{S_3 final} scales as $N^2$. To satisfy $\Gamma/H^4 \sim 1$, the nucleation temperature must decrease with $N$, see \cref{fig: S_3/T plots}. With greater supercooling, the critical bubble configuration decreases in radius, and the wall thickens. In this picture, the thin-wall approximation is rendered obsolete as the Euclidean action simply cannot be approximated by contributions from an infinitesimal wall and the bulk volume.

Even without the knowledge of the critical bubble profiles, one can still show that the thin-wall approximation breaks down at large $N$. The thin-wall formula for $S_3/T$ scales as $N^2$ in the large-$N$ limit, just like in the PLM; however, unlike that from the PLM, it is not a monotonically increasing function of temperature. As discussed above, it has a minimum between $T=0$ and $T=T_{\text{c}}$. Therefore, as $N$ is increased, a point will be reached at which $S_3/T$ (as a function of temperature) never falls low enough for the nucleation condition to be met. Here, it was found to be in the region around $N \simeq 170$. This logic is displayed in \cref{fig: S_3/T plots}, where the black horizontal line denotes the value of $S_3/T$ at which $\Gamma/H^4 \sim 1$. For $N = 6$, this line intersects the two $S_3/T$ curves at approximately the same temperature, whereas at $N = 20$ there is a larger discrepancy in the prediction for the degree of supercooling. If one were to plot $S_3/T$ for $N \gtrsim 170$, the thin-wall curve would never intersect the black line. The conclusion is that, while the thin-wall approximation may be valid for small values of $N$, the large-$N$ behaviour of GW parameters derived using this approximation should not be trusted. In \cref{sec:GW-spectrum B,sec:GW-spectrum}, we state more definitively the values of $N$ for which thin-wall is valid.

\begin{table}[t]
	\renewcommand{\arraystretch}{1.5}
	\begin{tabular}{|c|c|c|}
		\hline
		$N$ & $\tau_{\text{min}}$  & $\tau_{\text{n}}$ \\
		\hline
		3 & $0.99518^{+ 2 \cdot 10^{-5} \strut}_{-4.7 \cdot 10^{-4}}$ & $0.99931^{+5 \cdot 10^{-5} \strut}_{- 3 \cdot 10^{-5}}$ \\[1ex]
		4 & $0.99205^{+1.27 \cdot 10^{-3} \strut}_{-1.26 \cdot 10^{-3}}$ & $0.99774^{+ 2.7 \cdot 10^{-4} \strut}_{- 2.5 \cdot 10^{-4}}$ \\[1ex]
		5 & $0.98664^{+ 1.42 \cdot 10^{-3} \strut}_{- 1.46 \cdot 10^{-3}}$ & $0.99530^{+ 2.5 \cdot 10^{-4} \strut}_{- 2.5 \cdot 10^{-4}}$ \\[1ex]
		6 & $0.97219^{+ 2.61 \cdot 10^{-3} \strut}_{-4.92 \cdot 10^{-3}}$ & $0.99230^{+9.9 \cdot 10^{-4} \strut}_{- 8.1 \cdot 10^{-4}}$ \\[1ex]
		8 & $\tau_{\text{min}}^{N = 6}$ & $0.98855^{+9.7 \cdot 10^{-4} \strut}_{-7.0 \cdot 10^{-4}}$ \\[1ex]
		$ \rightarrow \infty$ & $\tau_{\text{min}}^{N = 6}$ & $\rightarrow \tau_{\text{min}}$ \\ 
		\hline
	\end{tabular}
	\caption{Minimum and nucleation temperatures, with associated upper- and lower-error bars, for $N = 3, 4, 5, 6, 8$, and $N \rightarrow \infty$ from the reconstructed PLM. The nucleation temperatures were computed using $T_{\text{c}} = 1$\,GeV, while the minimum temperatures are independent of this choice.}
	\label{tab: Nucleation and min temp table}
\end{table}

\begin{figure*}[t]
	\includegraphics[width=0.49\textwidth]{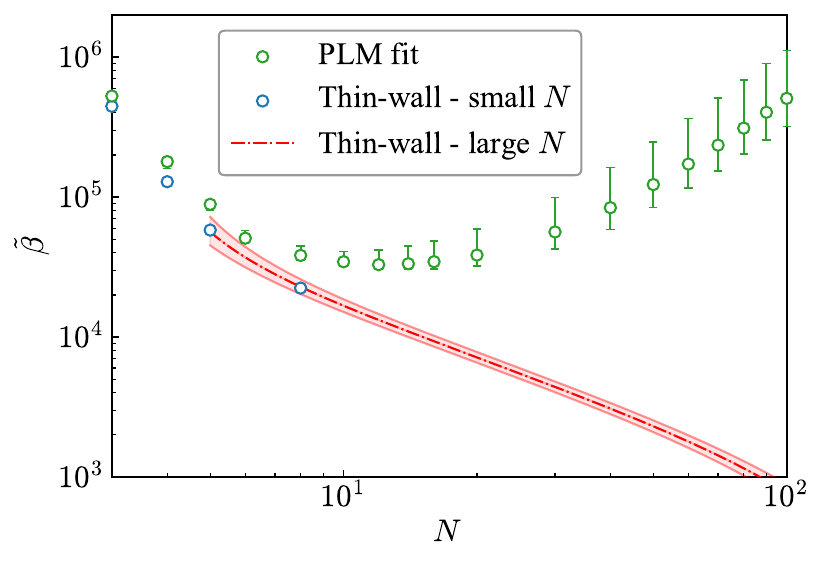}
	\hfill
	\includegraphics[width=0.49\textwidth]{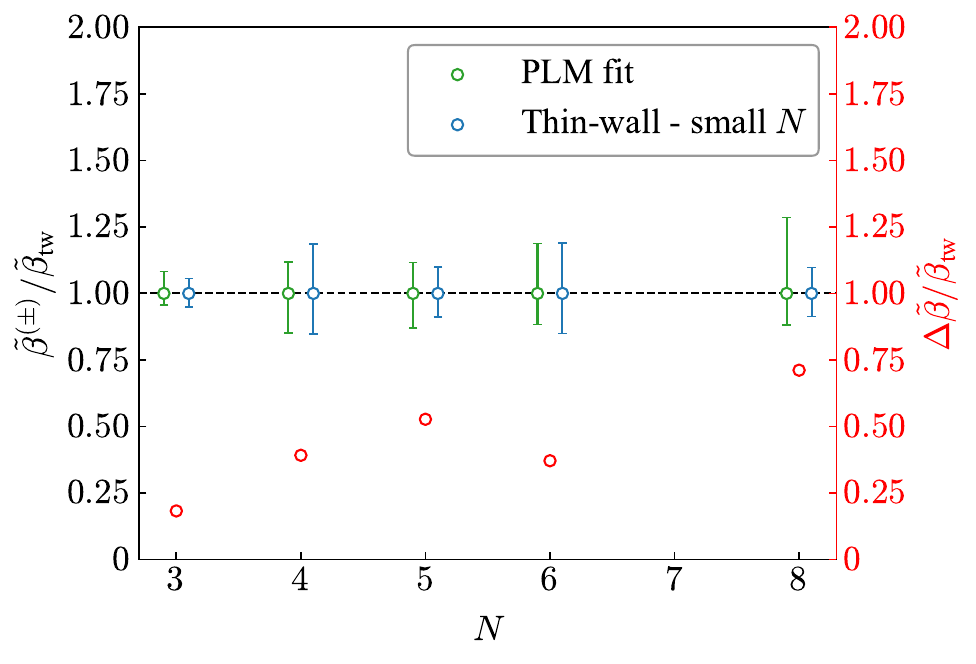}
	\caption{Inverse duration of the phase transition, in units of the Hubble parameter, plotted as a function of the number of colours. In the left-hand panel, the inverse duration is plotted at large $N$; the red curve is the result obtained through the thin-wall approximation in \cref{S_3 thin-wall large-N 2}, and that computed using the PLM fit is plotted in green. The right-hand panel displays the inverse duration at small $N$, where thin-wall predictions for $N = 3, 4, 5,$ and $8$ were made using infinite-volume-extrapolated values of the interface tension and latent heat instead of large-$N$ fits (and shown in the left-hand panel in blue). The left $y$-axis, $\tilde{\beta}^{(\pm)}/\tilde{\beta}_{\text{tw}}$, denotes the relative error in the inverse duration. The right $y$-axis gives the fractional difference between the inverse duration from the PLM and the thin-wall. The critical temperature was set to $T_{\text{c}}=1$ GeV in both cases.}
	\label{fig: Beta plots}
\end{figure*}

Using a critical temperature of $T_{\text{c}} = 1$ GeV, the PLM and thin-wall predictions for the nucleation temperature are plotted as a function of $N$ in \cref{fig: Nucleation temp plot}. For the reasons outlined above, at small values of $N$, the two approaches are in agreement. With increasing $N$, the results differ by a greater degree. A notable feature of the nucleation temperatures obtained from the PLM is that they quickly approach $\tau_{\text{min}}^{N = 6} \simeq 0.972$, denoted by the black dashed line. In the large-$N$ limit, the amount of supercooling is therefore bounded by this value and can be thought of as approximately constant. The constant nucleation temperature at large $N$ allows one to estimate the large-$N$ scaling of other GW parameters such as the inverse duration, $\tilde{\beta}$, and bubble wall velocity, $\xi_{\text{w}}$, as discussed in \cref{sec:GW-spectrum,sec:GW-spectrum B,sec:GW-spectrum C,sec:GW-spectrum D,sec:GW-spectrum E}. For small values of $N$, and as $N \rightarrow \infty$, these nucleation temperatures are tabulated alongside the corresponding minimum temperature of the deconfined phase in \cref{tab: Nucleation and min temp table}.

For each value of $N$ plotted in \cref{fig: Nucleation temp plot}, the nucleation temperature was also computed at a critical temperature of $T_{\text{c}} = 100$ GeV. The result was a marginal reduction in $\tau_{\text{n}}(N)$ (i.e.\ increased supercooling) when compared to that computed at $T_{\text{c}} = 1$ GeV across the entire range of $N$ used. This became increasingly negligible at large $N$ as $\tau_{\text{n}} \rightarrow \tau_{\text{min}}$.

At large $N$, the greatest source of uncertainty in the nucleation temperature stems from the rescaling procedure using the $SU(6)$ effective potential displayed in \cref{V scaling}, which sets the limit on supercooling to be $\tau_{\text{min}}^{N=6}$ in the large-$N$ limit. This rescaling works under the assumption that $N = 6$ is already large, such that one would not expect $\tau_{\text{min}}^N$ to decrease much with an increase in $N$. In the holographic model of \cite{Guersoy2009}, which is strictly valid for large $N$, a similar minimum temperature was observed to this $SU(6)$ effective potential. That being said, the lack of lattice data for $N > 8$ rendered this approximation a necessity to access the large-$N$ behaviour. The implications of this for each parameter entering the formula for the GW power spectrum in \cref{GW PS def} are discussed in detail in \cref{sec:GW-spectrum F,sec:Gw-spectrum-errors}. Notably, the maximal degree of supercooling for $N \geq 3$ was estimated in \cite{Agrawal:2025xul}, where a slightly larger value was obtained. 

\subsection{Inverse Duration, $\tilde{\beta}$}
\label{sec:GW-spectrum B}

Next, we examine the phase transition inverse duration, $\beta$. With the bubble nucleation rate defined in \cref{Nucleation rate def}, the inverse duration is written as \cite{Hindmarsh2021},
\begin{equation} \label{Beta def}
	\beta \equiv \left. \frac{\mathrm d}{\mathrm dt} \log \Gamma (t) \right|_{t = t_{\text{n}}} ,
\end{equation}
where the nucleation time, $t_{\text{n}}$, is taken to be the time of GW production. For a sufficiently fast phase transition (i.e.\ $\beta \gg H$), the nucleation rate can be approximated by
\begin{equation} \label{Nucleation rate with beta}
	\Gamma(t) \simeq \Gamma(t_{\text{n}}) e^{\beta(t - t_{\text{n}})}.
\end{equation}
A Taylor expansion of $S_3$ in \cref{Nucleation rate def} around $t = t_{\text{n}}$ allows for the identification
\begin{equation} \label{Beta with S_3}
	\beta = \left. -\frac{\mathrm d}{\mathrm dt} \frac{S_3(T)}{T} \right|_{t = t_{\text{n}}}.
\end{equation}
In practice, the inverse duration is made dimensionless by dividing by $H$. Using that $\mathrm dT/\mathrm dt = -HT$,
\begin{equation} \label{Dimensionless beta}
	\tilde{\beta} \equiv \frac{\beta}{H(T_{\text{n}})} = \left. T \frac{\mathrm d}{\mathrm dT} \frac{S_3(T)}{T} \right|_{T = T_{\text{n}}}.
\end{equation} 
From the Euclidean action in the thin-wall approximation \cref{S_3 thin-wall large-N 2}, the inverse duration can be computed analytically, whereas that using the PLM must be obtained numerically. In \cref{fig: Beta plots}, the dimensionless inverse duration computed using both the thin-wall approximation and the PLM is plotted as a function of the number of colours at a critical temperature of $T_{\text{c}} = 1$ GeV. For small values of $N$ $\left(\leq 6\right)$, the two approaches are in good agreement, as is demonstrated in the right-hand panel; this can be attributed both to the matching procedure described in \cref{sec:PLM B} and to the relatively small degree of supercooling. Indeed, as $N$ increases from $3$ to $8$, the greater discrepancy in supercooling begins to take hold and the estimates of $\tilde{\beta}$ disagree more strongly. In the left-hand panel of \cref{fig: Beta plots}, the inverse duration from the PLM is plotted for a larger range of $N$, and is accompanied by the prediction made using the thin-wall formula of \cref{S_3 thin-wall large-N 2}.

At large $N$, the inverse durations computed in the thin-wall approximation and using the PLM are in disagreement. Although the interface tension and latent heat derived from the Polyakov loop potential exactly agree with the lattice large-$N$ fits, the discrepancy in the observed supercooling of \cref{fig: Nucleation temp plot} takes over and the predictions for $\tilde{\beta}$ begin to mutually diverge. For the reasons outlined in \cref{sec:GW-spectrum A}, the thin-wall computation should not be trusted in this regime. A striking result in the PLM is the observation of a turning point in $\tilde{\beta}$ at $N \simeq 12$ (notably, the same behaviour was observed in \cite{Huang2021} at $N \simeq 11$ for different reasons). As discussed in \cref{sec:GW-spectrum}, this feature has a significant effect on the behaviour of the peak amplitude and peak frequency of the GW power spectrum as a function of $N$. It should be noted that the asymmetry in the error bars stems from the corresponding asymmetric error in the nucleation temperature.

In the large-$N$ regime, one would naively expect that the inverse duration scales as $N^2$. The reason is that the effective potential scales with $N^2$, see \cref{V scaling}, and so does the three-dimensional Euclidean action $S_3$. Other $N$-dependences can arise from the $N$-dependence of the nucleation temperature $\tau_\text{n}$, which is, however, approximately constant at large $N$. In rough agreement with this expectation, we find numerically that $\tilde{\beta} \propto N^{2.12}$ for $N \geq 60$. The small discrepancy arises due to the nucleation temperature being only approximately constant. The large-$N$ power-law fit parameter of the inverse duration is displayed in \cref{tab: large-N GW peak amp} in \cref{sec:peak-amplitude-and-frequency} for $T_{\text{c}} = 1$ GeV and $T_{\text{c}} = 100$ GeV, where we discuss the implications on the GW power spectrum peak amplitude.

The presence of a turning point in $\tilde\beta$ can be explained by two competing effects in the small- and large-$N$ regimes. At large $N$, we identified above that $\tilde\beta$ scales approximately as $\tilde\beta \propto N^2$ because the limit on supercooling is being approached and the system is dominated by the number of degrees of freedom, $n_{\text{dof}} \propto N^2$. At small $N$ the thin-wall approximation is valid. In this limit, the inverse duration can be described by the relation $\tilde\beta \propto L/\sigma^{3/2}$ \cite{Reichert:2021cvs}, and substituting the large-$N$ formulae \cref{Large-N fits new} ($N \geq 5$) we find $\tilde\beta \propto 1/N$. Note that while for $N < 5$ the interface tension and latent heat do not follow their large-$N$ behaviour, we still observe inverse scaling of $\tilde\beta$ with $N$. The turning point appears when these two scaling effects cancel out, and can therefore be attributed to the nucleation temperature approaching $T_{\text{min}}$.

The inverse duration of the phase transition was also computed at a critical temperature of $T_{\text{c}} = 100$ GeV for each $N$. The effect was to slightly scale down $\tilde{\beta}$ by a factor which became increasingly negligible as the nucleation temperature approached $\tau_{\text{min}}$ (i.e.\ as $N$ became large). The turning point in $\tilde{\beta}$ remained at $N \simeq 12$.

Looking ahead, we will see that the inverse duration is the dominating quantity in the prediction of the GW power spectrum. Across the entire range of $N$ it has a large magnitude, which generally suppresses the spectrum, and its shape as a function of $N$ greatly influences the shape of the peak amplitude with $N$. The reason for such a large $\tilde{\beta}$ can be attributed to the rapidly changing effective potential with temperature, which in turn causes the Euclidean action to vary steeply. The physical picture is that below the critical temperature, a large number of very small true vacuum bubbles nucleate in the gluon plasma ($\beta \sim 1/R$). These bubbles expand, collide, and generate sound waves in this plasma, but due to the small radii, we expect the resulting GW signal to be weak. 

\subsection{Bubble Wall Velocity, $\xi_{\text{\normalfont{w}}}$}
\label{sec:GW-spectrum C}
The non-perturbative nature of first-order confinement phase transitions renders the computation of the bubble wall velocity a formidable task. Indeed, it has often been left as a free parameter \cite{Helmboldt:2019pan, Kang:2021epo, Reichert:2021cvs, Huang2021,Morgante2023, Halverson2021, Pasechnik2024, Ares:2020lbt}, introducing a significant degree of uncertainty in the resulting GW power spectrum. More recently, progress has been made in the determination of the bubble wall speed in strongly coupled transitions, particularly through the use of holographic models \cite{Bea2021, Bigazzi2021, Bea2022, Janik2022}. In this work, we follow \cite{SanchezGaritaonandia2024}, which develops a framework to estimate the wall speed in theories that experience a large change in the number of degrees of freedom across the phase transition, in a model-independent way. The $SU(N)$ Yang-Mills theory lies within this regime, particularly in the large-$N$ limit. 

\begin{figure*}[t]
	\includegraphics[width=0.48\textwidth]{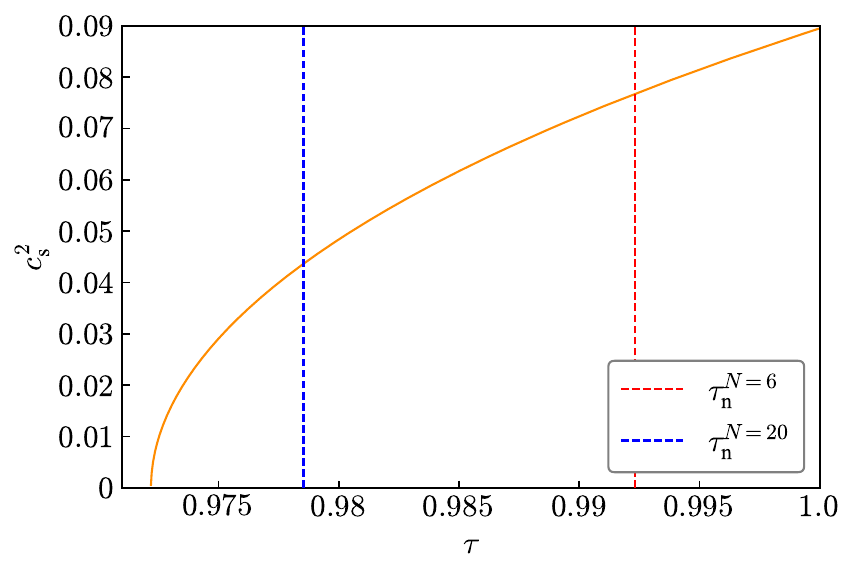}
	\hfill
	\includegraphics[width=0.49\textwidth]{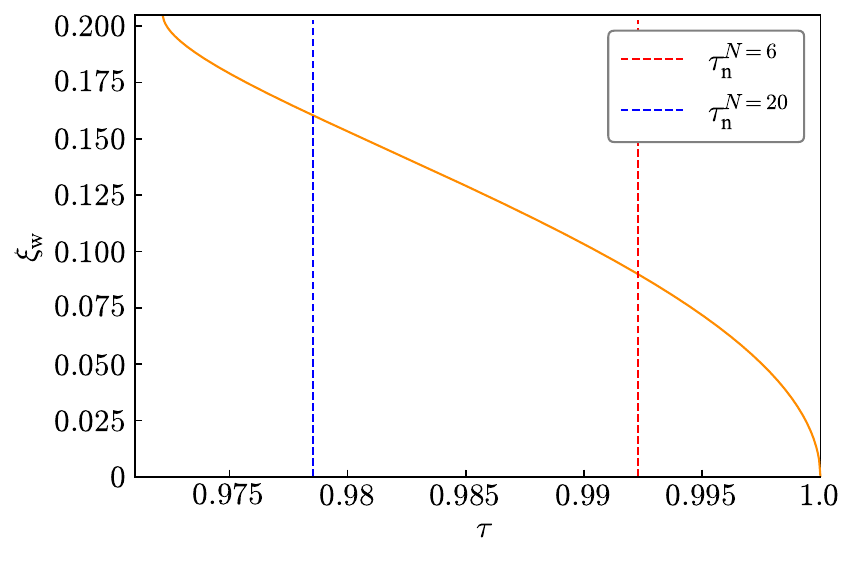}
	\caption{Squared sound speed of the high-enthalpy phase (left) and bubble wall velocity (right) as a function of the temperature ($\tau_{\text{min}} \leq \tau \leq 1$).The curves were obtained for $N \geq 6$ and $T_{\text{c}} = 1$\,GeV. The nucleation temperatures for $N=6$ and $N=20$ are indicated with vertical dashed lines.}
	\label{fig: Sound and Wall speed plots}
\end{figure*}

A review of the hydrodynamics related to computing the bubble wall speed can be found in \cite{Espinosa2010}. The imposition of conservation of energy-momentum across the phase transition boundary allows one to derive the following matching conditions (under the assumption that the outward-moving wall is infinitesimally thin, which is a good approximation in most cases):
\begin{align} \label{Matching conditions 1}
	w_+ v^2_+ \gamma^2_+ + p_+ &= w_- v^2_- \gamma^2_- + p_-\,, \notag \\
	w_+ v^2_+ \gamma^2_+ &= w_- v^2_- \gamma^2_-\,,
\end{align} 
where a $+$ ($-$) denotes evaluation just ahead of (behind) the bubble wall, $w_{\pm} = T \partial p_{\pm}/\partial T$ is the enthalpy, $v_{\pm}$ represents the fluid velocity in the wall frame, and $\gamma_{\pm} = 1/\sqrt{1 - v_{\pm}^2}$. The following can then be obtained:
\begin{align} \label{Matching conditions 2}
	v_+v_- &= \frac{p_+ - p_-}{e_+ - e_-}\,,
	&& \text{and}&
	\frac{v_+}{v_-} &= \frac{e_- + p_+}{e_+ + p_-}\,.
\end{align}
The hydrodynamic equations admit three classes of solutions:
\begin{itemize}
	\item[(I)] Deflagration: the fluid is at rest behind the wall ($v_- = \xi_{\text{w}}$). A shock is present, ahead of which the temperature is equal to the nucleation temperature ($T^+_{\text{sh}} = T_{\text{n}}$). 
	\item[(II)] Detonation: the fluid is at rest in front of the wall ($v_+ = \xi_{\text{w}}$). No shock is present, and the temperature in front of the wall is the nucleation temperature ($T_{+} = T_{\text{n}}$).
	\item[(III)] Hybrid: a superposition of the above solutions. The fluid velocity behind the wall is equal to the speed of sound in the low-temperature phase ($v_- = c_{\text{s}-}$). A shock is present, ahead of which the temperature is equal to the nucleation temperature ($T^+_{\text{sh}} = T_{\text{n}}$).
\end{itemize}
To proceed, one must find a way to relate $v_+$ to $v_-$, or equivalently $T_+$ to $T_-$. In the presence of a plasma, the bubble wall quickly reaches a terminal velocity; to obtain the relation between $\pm$ quantities in full generality requires an analysis of the friction forces on the wall. As was alluded to above, this is a challenging task, particularly for strongly coupled theories. Another approach is to compute an upper- and lower-bound on the wall velocity, and treat it as a free parameter in this band. For a discussion of these limits and an application thereof to the QCD phase transition, see \cite{Ai2025} and \cite{Cline:2025bwe}, respectively.

In \cite{SanchezGaritaonandia2024}, a large jump in enthalpy is imposed by the suppression of quantities in the low-enthalpy (confined) phase by $N^2$ in the matching conditions of \cref{Matching conditions 1,Matching conditions 2}. Through considerations of the scaling of these matching conditions in the large-$N$ limit, the following constraints were derived:
\begin{align} \label{Large-enthalpy conditions}
	T_+ &= T_{\text{c}}\,, 
	&&\text{and}&
	v_+ &= 0\,.
\end{align}
The above conditions agree with the results of the holographic simulations in \cite{Bea2021, Janik2022}, and allow one to forgo the use of bounds on the wall velocity. Clearly, the second constraint implies the exclusion of detonations. Using \cref{Large-enthalpy conditions}, the hydrodynamics equations were solved in \cite{SanchezGaritaonandia2024} for the strongly coupled holographic model presented in \cite{Guersoy2009} to obtain the bubble wall speed as a function of the nucleation temperature. Importantly, this holographic model exhibits similar behaviour to the PLM regarding the limit on supercooling, and one would therefore expect similar predictions for the sound speed and bubble wall speed in this work. Both models make thermodynamic predictions only of the high-enthalpy phase, which here is the deconfined phase, but the large-$N$ conditions \cref{Large-enthalpy conditions} allow the computation of the wall speed to proceed without any knowledge of the fluid variables behind the bubble wall.

Thermodynamics tells us that the sound speed is given in terms of the pressure and energy density, $e$, as
\begin{align} \label{Sound speed}
	c_{\text{s}}^2 \equiv \frac{\partial p}{\partial e} &= \frac{\partial p/ \partial T}{\partial e/\partial T}\,,
	&
	e &= T \frac{\partial p}{\partial T} - p \,,
\end{align}
which we evaluate at $\ell_0$ for the deconfined phase, i.e.\ $p = p_+ \simeq -V_{\text{eff}}(\ell = \ell_0, T)$. More precisely, the PLM only gives us access to $\Delta p \simeq - \left( V_{\text{eff}} (\ell = \ell_0, T) - V_{\text{eff}} (\ell = 0, T) \right)$ and it implicitly sets $p_- = 0$. Under the assumption that the absolute pressure $p_+$ is shifted by a constant, this drops out of the temperature derivatives in \cref{Sound speed}. Using \cref{Sound speed}, the sound speed was computed from the reconstructed Polyakov loop potentials, the result of which is plotted on the left-hand side of \cref{fig: Sound and Wall speed plots} for $N \geq 6$. Qualitatively, the sound speed curve is in good agreement with that obtained from the dark $SU(N)$ holographic model of \cite{Guersoy2009}, differing by a small amount at the critical temperature. This stems from a slight discrepancy in the prediction for $\tau_{\text{min}}$. At large $T$ compared to the critical temperature, the speed of sound in the deconfined phase approaches its conformal value of $c_{\text{s}} \simeq 1/\sqrt{3}$ as required. 

Employing the large-enthalpy-jump conditions \cref{Large-enthalpy conditions}, the hydrodynamics of the confinement transition in the PLM were solved. The bubble wall speed for $N \geq 6$ is plotted on the right-hand side of \cref{fig: Sound and Wall speed plots} as a function of temperature ($T_{\text{min}} \leq T \leq T_{\text{c}}$). As expected, given the similarities in the predicted sound speed and the limit on supercooling between the PLM and the holographic model of \cite{Guersoy2009}, the bubble wall velocity curves are in good qualitative agreement. Here, a maximum wall velocity of $\xi_{\text{w}} \simeq 0.20$ is observed, which can be thought of as the limiting value of $\xi_{\text{w}}$ that is approached at large $N$. Using the model in \cite{Guersoy2009}, a maximum value of $\xi_{\text{w}} \simeq 0.25$ was obtained in \cite{SanchezGaritaonandia2024}, which can be explained by the slightly lower minimum temperature when compared to the PLM. Note that the wall velocity might be smaller if the $SU(N)$ dark sector is coupled to the Standard Model, such that the large latent heat released in the transition can reheat the Universe outside the true vacuum bubble \cite{Asadi:2021pwo, Gouttenoire:2023roe}.

Strictly speaking, the right-hand-side plot of \cref{fig: Sound and Wall speed plots} actually displays the solution for $v_-$ as a function of temperature below the critical, which only equates to the wall speed for a deflagration \cite{Espinosa2010}. In this case, the fluid is at rest inside the bubble such that $v_- = \xi_{\text{w}}$. While the large-enthalpy limit excludes detonations, made explicit by the conditions of \cref{Large-enthalpy conditions}, hybrid solutions remain allowed. For hybrids, $v_- = c_{\text{s}-}$ and $c_{\text{s}-} < \xi_{\text{w}} < v_{\text{CJ}}$, where $v_{\text{CJ}}$ is the Chapman-Jouguet velocity; the minimum wall speed of a detonation. With knowledge of the sound speed in the confined phase, one should then compute the wall velocity through a shooting method. As is stated above, the PLM gives us no information on the sound speed in the low-enthalpy phase. We therefore assume that it quickly returns to that of a relativistic ideal gas with $c_{\text{s}-} = 1/\sqrt{3}$, implying deflagration solutions only. Under this assumption, $v_- = \xi_{\text{w}}$ and the right-hand side of \cref{fig: Sound and Wall speed plots} is indeed the wall velocity. 

Notably, in the holographic study of the bubble wall velocity in \cite{Bea2021}, the sound speed in the low-enthalpy phase was successfully computed. It was found to intersect that of the deconfined phase at $T = T_{\text{c}}$ and quickly tend towards the conformal value with decreasing temperature. Given a suitable equation of state describing the low-enthalpy phase, it is reasonable to assume that similar behaviour would be observed using the PLM. In this case, the confined sound speed, $c_{s-}$, would be bounded by $c^2_{\text{s}+}(T=T_{\text{c}}) \simeq 0.09 \leq c^2_{\text{s}-} \leq 1/3$, admitting only deflagrations as before (since $c_{\text{s}-} \geq \sqrt{0.09} > v_- = \xi_{\text{w}}$ by \cref{fig: Sound and Wall speed plots}). 

\begin{figure}[t]
	\includegraphics[width=\linewidth]{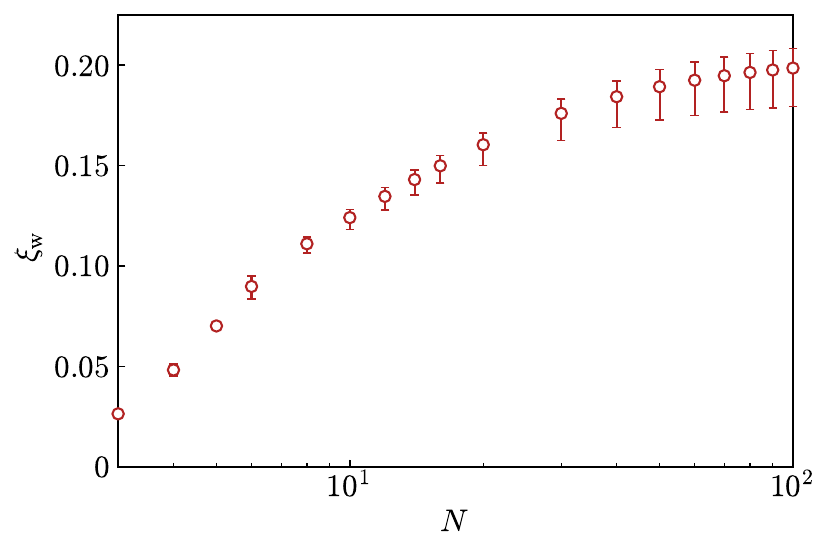}
	\caption{Bubble wall velocity evaluated at the nucleation temperature and plotted as a function of $N$ for $T_{\text{c}} = 1$ GeV.}
	\label{fig: Large-N wall and sound speed plot}
\end{figure} 

We would like to point out a potential source of confusion. The curves of the sound speed in the deconfined phase and the wall velocity intersect at some temperature, see \cref{fig: Sound and Wall speed plots}. Naively, this would imply that the bubble wall moves outwards with a speed greater than the sound speed of the plasma, such that in this regime we actually find detonations. This would be in contradiction to the second large-enthalpy-jump condition in \cref{Large-enthalpy conditions} used in the computation. The contradiction is resolved by recalling the presence of a shock front in which reheating of the plasma occurs, and that the first of the large-enthalpy-jump conditions in \cref{Large-enthalpy conditions} implies this plasma is reheated up to the critical temperature in front of the bubble wall. Therefore, the bubble wall interacts with a plasma of sound speed $c_{s+}(T_{\text{c}})$ and detonations are indeed excluded, given that the wall velocities we compute never exceed this value.

These considerations lead us to the predictions of the bubble wall velocity from the PLM as a function of $N$ displayed in \cref{fig: Large-N wall and sound speed plot}. The wall velocity is monotonically increasing and approaches $\xi_{\text{w}} \simeq 0.20$ at large $N$. The small value of $\xi_w$ at small $N$ provides an additional suppression of the GW power spectrum at small $N$.

\subsection{Phase Transition Strength Parameter, $\alpha$}
\label{sec:GW-spectrum D}

The total energy available to be converted into GWs is given by the change in the trace of the energy-momentum tensor across the phase transition. In this work, we define the dimensionless phase transition strength parameter, $\alpha$, in terms of the trace anomaly, $\theta$, which is proportional to the trace of the energy-momentum tensor. In terms of the pressure and energy density, the trace anomaly is given by
\begin{equation} \label{Theta def}
	\theta_{\pm} = \frac{1}{4} \left( e_{\pm} - 3p_{\pm} \right).
\end{equation}
Defining $\Delta \theta = \theta_+ - \theta_-$, which therefore encodes the energy available for conversion to shear stress, the strength parameter is written as
\begin{equation} \label{Strength param def}
	\alpha = \left. \frac{4}{3} \frac{\Delta \theta}{w_+} \right|_{T = T_{\text{n}}},
\end{equation}
where $w_+$ quantifies the degrees of freedom in the deconfined phase.

\begin{figure}[t]
	\includegraphics[width=\linewidth]{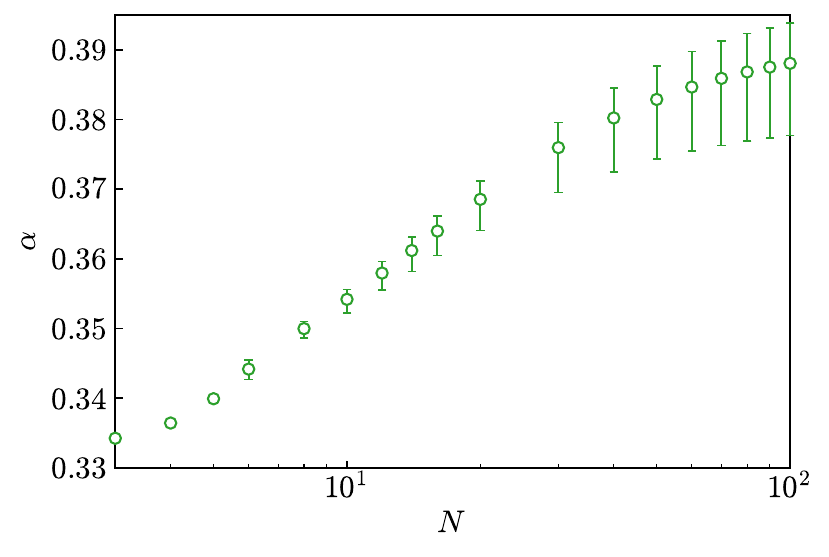}
	\caption{Phase transition strength parameter, calculated from \cref{Strength param def} at $T_{\text{c}} = 1$ GeV, plotted as a function of $N$.}
	\label{fig: Strength param plot}
\end{figure}

To access the pressure, we use $p_{\pm} \simeq - V_{\text{eff}}(\ell_{\pm}, T) \equiv -V_{\text{eff} \pm}$, where $\ell_{\pm}$ denotes $\ell = \ell_0$ and $\ell = 0$ respectively. The potential is normalised to $V_{\text{eff}}(\ell = 0, T) = 0$, which allows one to express \cref{Theta def} in terms of quantities evaluated in the deconfined phase:
\begin{equation} \label{Strength param potential}
	\alpha = \left. \frac{1}{3} \frac{T \partial V_{\text{eff} +}/\partial T - 4 V_{\text{eff} +}}{T \partial V_{\text{eff} +}/\partial T} \right|_{T = T_{\text{n}}}.
\end{equation}
For transitions which occur at $T_{\text{n}} \simeq T_{\text{c}}$, such that $V_{\text{eff}+} \simeq 0$, we find $\alpha \simeq 1/3$, see also \cite{Huang2021, Halverson2021, Reichert:2021cvs, Morgante2023}. Therefore, given the nucleation temperatures plotted in \cref{fig: Nucleation temp plot}, one expects $\alpha \simeq 1/3$ for small values of $N$, and that at large $N$ the strength parameter should asymptotically approach some value, given that the nucleation temperature remains approximately constant ($\simeq \tau_{\text{min}}^{N=6}$). We show the transition strength parameter in \cref{fig: Strength param plot}, and in the large-$N$ limit it approaches $\alpha \simeq 0.39$.

\subsection{Efficiency Factor, $\kappa$}
\label{sec:GW-spectrum E}
The final quantity to be defined, which enters the computation of the GW power spectrum \cref{GW PS def}, is the efficiency factor. In our definition, we split up $\kappa$ into 
\begin{align} \label{eq:def-kappa}
	\kappa = \sqrt{\tau_{\text{sw}}}\, \kappa_{\text{sw}}\,,
\end{align}
where $\tau_{\text{sw}}$ encodes the lifetime of the source, and $\kappa_{\text{sw}}$ is the usual efficiency factor. The efficiency factor is defined as the ratio of the total kinetic energy in the plasma to the vacuum energy, and is therefore a measure of the total energy fraction which can go into GW production; the rest is used to increase the thermal energy of the system. Its definition reads
\begin{align} \label{Kappa def}
	\kappa_{\text{sw}} = \frac{3}{\epsilon \,\xi_{\text{w}}^3} \int \! \mathrm d\xi \, w(\xi) v(\xi)^2 \gamma(\xi)^2 \xi^2\,,
\end{align}
where $\epsilon$ is the vacuum energy released in the transition, $\gamma$ is the Lorentz factor, $w$ is the enthalpy, and $\xi$ is a coordinate in the frame of the bubble centre. To obtain the efficiency factor in full generality requires the fluid velocity profile, $v(\xi)$, which can be computed by integrating the fluid equations, as is discussed in \cite{Espinosa2010}, and one should include the temperature-dependent sound speed. This is particularly relevant where it deviates strongly from the conformal value, such as in the present scenario.

In situations where the hydrodynamics is not precisely known, it is simpler to use fitted formulae for $\kappa_{\text{sw}}$ from numerical simulations \cite{Espinosa2010}. Relevant in this study, given the wall velocity is bounded to be a deflagration, are the following:
\begin{align} \label{Kappa fits}
	\kappa_{\text{A}} &\simeq \xi_{\text{w}}^{6/5} \frac{6.9 \alpha}{1.36 - 0.037 \sqrt{\alpha} + \alpha}\,, & \xi_{\text{w}} &\ll c_{\text{s}}\,, \notag\\[1ex]
	\kappa_{\text{A} \rightarrow \text{B}} &\simeq \frac{c_{\text{s}}^{11/5} \kappa_{\text{A}} \kappa_{\text{B}}}{\left(c_{\text{s}}^{11/5} - \xi_{\text{w}}^{11/5}\right) \kappa_{\text{B}} + \xi_{\text{w}} c_{\text{s}}^{6/5} \kappa_{\text{A}}}\,, & \xi_{\text{w}} &\lesssim c_{\text{s}}\,,\notag\\[1ex]
	\kappa_{\text{B}} &\simeq \frac{\alpha^{2/5}}{0.017 + (0.997 + \alpha)^{2/5}}\,, &\xi_{\text{w}}  &= c_{\text{s}}\,, 
\end{align}
where $c_{\text{s}} \equiv c_{\text{s}+}(T_+ = T_{\text{c}})$ is understood. Given the assumption of a constant sound speed, the use of the above formulae should be taken as a conservative estimate of $\kappa_{\text{sw}}$. We expect the efficiency factor in a full calculation to be larger and we reserve this investigation for a future study.

In the computation of the GW power spectrum, we choose the appropriate fit for the efficiency factor for each value of $N$ depending on the value of the wall velocity displayed in \cref{fig: Large-N wall and sound speed plot}. Specifically, we choose: (i) $\xi_{\text{w}} < 0.25 \, c_{\text{s}} \implies \kappa_{\text{sw}} = \kappa_{\text{A}}$, and (ii) $0.25 \, c_{\text{s}} \leq \xi_{\text{w}} < c_{\text{s}} \implies \kappa_{\text{sw}} = \kappa_{\text{A} \rightarrow \text{B}}$. Since the wall velocity is lower at small $N$, the efficiency factor $\kappa_{\text{sw}}$ further suppresses the GW spectrum at small $N$.

To determine the GW power spectrum from an acoustic source, one must also take into account the lifetime of this source \cite{Ellis2020}. We introduce a factor of $\sqrt{\tau_{\text{sw}}}$, the length of the sound wave production period, in the definition of $\kappa$ \cref{eq:def-kappa}. The amplitude of the power spectrum therefore grows linearly with the source duration as $h^2 \Omega_{\text{GW}}^{\text{peak}} \propto \kappa^2$. In an expanding universe, this factor is given by \cite{Guo2021}
\begin{equation} \label{Source duration}
	\tau_{\text{sw}} = 1 - \left[ 1 + 2 \frac{(8 \pi)^{1/3} \xi_{\text{w}}}{\tilde{\beta} U_f} \right]^{-1/2},
\end{equation}
which acts to significantly suppress the peak amplitude of the spectrum. In \cref{Source duration}, $U_f$ denotes the root-mean-squared fluid velocity, which takes the following form:
\begin{equation} \label{RMS fluid speed}
	U_f^2 \simeq \frac{3}{\xi_{\text{w}}(1 + \alpha)} \int_{c_{\text{s}}}^{\xi_{\text{w}}} \mathrm d \xi \frac{\xi^2 v(\xi)^2}{1 - v(\xi)} \simeq \frac{3}{4} \frac{\alpha}{1 + \alpha} \kappa_{\text{sw}}\,.
\end{equation}
Notably, in the simulations of \cite{Caprini:2024gyk} it was found that the GW amplitude continued to grow after $\tau_{\text{sw}}$ (which strictly speaking denotes the timescale over which non-linearities develop, but can be thought of as that beyond which efficient sound wave generation ceases), albeit much slower than linearly. In this sense, the introduction of $\tau_{\text{sw}}$ into the formula for the GW power spectrum in \cref{GW PS def}, which enforces that the source is abruptly turned off at this point, is a lower-bound on the timescale of GW production from sound waves. However, given the slowly-growing amplitude after $\tau_{\text{sw}}$ with time, this likely remains a reasonable estimate. 

\begin{figure*}[t]
	\includegraphics[width=0.49\textwidth]{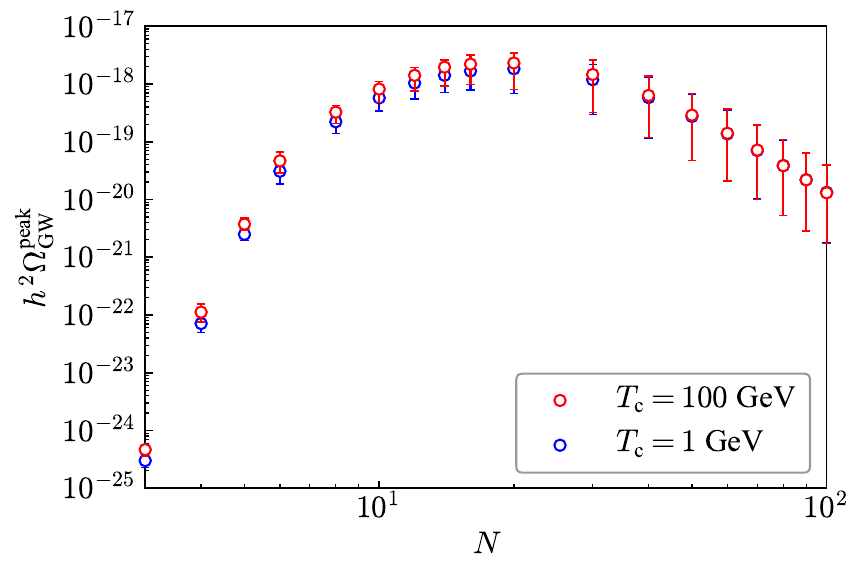}
	\hfill
	\includegraphics[width=0.48\textwidth]{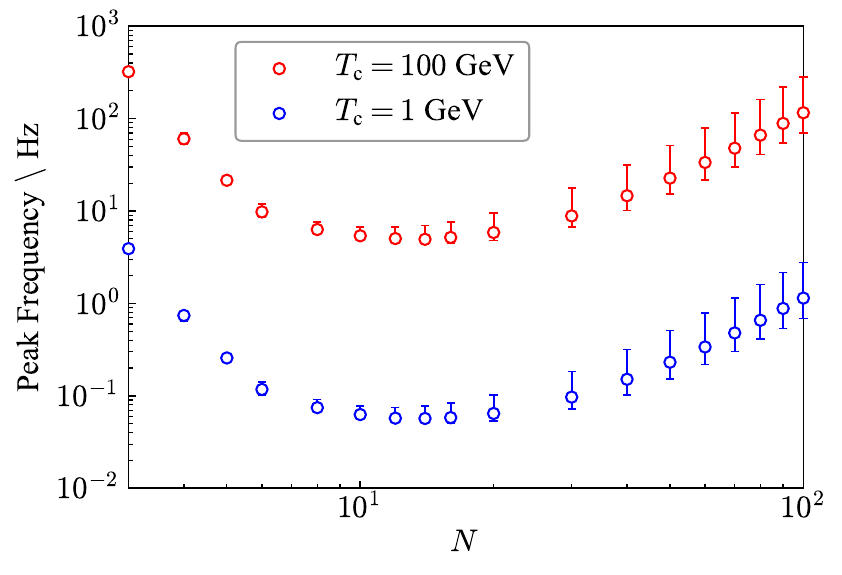}
	\caption{Peak amplitude (left) and peak frequency (right) of the computed stochastic GW background, plotted as a function of the number of colours, shown for both $T_{\text{c}} = 1$ GeV and $T_{\text{c}} = 100$ GeV.}
	\label{fig: Large-N GW plots}
\end{figure*}

\subsection{GW Parameters at Large $N$}
\label{sec:GW-spectrum F}
Above, we have displayed the results for each parameter that enters the formula of the GW power spectrum \cref{GW PS def}. Here, we discuss the observed large-$N$ scaling of these parameters and the uncertainties in the large-$N$ limit.

The PLM fitting is based upon lattice data at and around the critical temperature. Consequently, properties further away from the critical temperature have a larger uncertainty. As such, the minimum temperature $\tau_{\text{min}}$ where the potential barrier disappears has a rather large uncertainty in our fitting procedure. Additionally, we use the rescaled $N=6$ data to determine the large-$N$ Polyakov loop potential and hence $\tau_{\text{min}}^{N=6} \simeq \tau_{\text{min}}^{N \rightarrow \infty}$. While we expect the minimum temperature of the deconfined phase in $SU(N)$ to tend to a constant in the large-$N$ limit, it could be lower than that for $SU(6)$. This leads to an additional uncertainty in the GW parameters at large $N$, which is not quantified in their error bars. We discuss the implications of a lower $\tau_{\text{min}}^{N \rightarrow \infty}$ in the following.

The inverse duration displayed in \cref{fig: Beta plots} scales as $N^{2.12}$ for $N \geq 60$ as discussed in \cref{sec:GW-spectrum B}. If $\tau_{\text{min}}^{N}$ does continue to decrease beyond that at $N = 6$, the greater degree of supercooling would cause $S_3/T$ to vary less steeply with temperature than is observed for $N = 6$ in \cref{fig: S_3/T plots}, leading $\tilde{\beta}$ to be slightly suppressed for larger values of $N$. The turning point in the inverse duration would then be pushed to a larger $N$, and so too would the onset of the approximate $N^2$ scaling. In the case of a lower $\tau_{\text{min}}^{N \rightarrow \infty}$, we therefore would expect $\tilde{\beta}$ to be shifted downwards for $N \geq 6$.

The wall velocity in the PLM approaches $\xi_{\text{w}} \simeq 0.20$ at large $N$, see \cref{fig: Large-N wall and sound speed plot}. In \cref{sec:GW-spectrum C}, we argued that this is due to $\tau_{\text{min}}$ approaching a constant at large $N$, and the value of $\tau_{\text{min}}^{N \rightarrow \infty}$ provides the largest uncertainty on the wall velocity. If $\tau_{\text{min}}$ saturates at a smaller value, then the wall velocity approaches a larger value, and the shape of the curve would be similar to that found here. Due to the comparison with \cite{SanchezGaritaonandia2024}, where a maximum value of $\xi_{\text{w}} \simeq 0.25$ was found, we are confident that we have presented a reasonable estimate of the wall velocity at large $N$.

The phase transition strength parameter plotted in \cref{fig: Strength param plot} tends towards a constant value in the large-$N$ limit. This is explained by the presence of a minimum temperature of the deconfined phase, which limits the degree of supercooling. If $\tau_{\text{min}}^{N \rightarrow \infty}$ was lower than that at $N = 6$, the strength parameter would approach a slightly larger value than was found here. This would have a minimal effect.

Overall, the GW parameter with the largest uncertainty at large $N$ is the inverse duration, which in turn is dominated by the uncertainty on the minimum temperature $\tau_{\text{min}}$ at large $N$. While we are confident that we have correctly captured the large-$N$ scaling, the value at which the scaling sets in could be at slightly larger $N$.

\section{GW Spectrum }
\label{sec:GW-spectrum}
We have displayed and discussed in \cref{sec:GW-spectrum params} the results for each parameter contributing to the GW power spectrum \cref{GW PS def}. In this section, we discuss the features of the resulting GW spectra and quantify their detectability at future GW observatories through the signal-to-noise ratio.

\subsection{Peak Amplitude and Frequency}
\label{sec:peak-amplitude-and-frequency}
We display the peak amplitude of the GW power spectrum, $h^2 \Omega_{\text{GW}}^{\text{peak}}$, as a function of $N$ in the left panel of \cref{fig: Large-N GW plots}. The peak amplitude has a maximum at $N \simeq 20$ but has a small value overall, $\mathcal{O}(10^{-24})$ for $N=3$ and  $\mathcal{O}(10^{-18})$ at the maximum.	Qualitatively, the peak amplitude of the GW power spectrum is a mirror image of the $\tilde{\beta}$ curve in \cref{fig: Beta plots}, which is the most important parameter for this quantity. Note that the extremum is shifted slightly to larger $N$, from $N \simeq 12$ in the $\tilde{\beta}$ curve, to $N \simeq 20$ in $h^2 \Omega_{\text{GW}}^{\text{peak}}$. This is due to the wall velocity, the strength parameter, and the efficiency factor increasing steadily with $N$.

In \cref{fig: Large-N GW plots}, the peak amplitude is plotted for both $T_{\text{c}} = 1$ GeV and $T_{\text{c}} = 100$ GeV. The slight variation in $h^2 \Omega_{\text{GW}}^{\text{peak}}$ between each $T_{\text{c}}$ is as a result of the slightly different nucleation temperatures ($\tau_{\text{n}} = T_{\text{n}}/T_{\text{c}}$) obtained from \cref{Nucleation rate def}. This effect is minimal due to the exponential suppression of the bubble nucleation rate \cref{Nucleation rate def} by $S_3/T$ and Planck suppression of $T_{\text{c}}$. The differences between $\tau_{\text{n}}$, and therefore the peak amplitudes, for each critical temperature generally shrink with $N$. Again, this is due to the fact that the dimensionless nucleation temperature approaches $\tau_{\text{min}}$ irrespective of the critical temperature used in the computation.

At large $N$, the peak amplitude follows an approximate power law. We fitted our results for $N \geq 60$ to
\begin{align} \label{Large-N GW peak amp func}
	f(N) = A \, N^B\,.
\end{align}
The fitting results for $T_{\text{c}} = 1$ GeV and $T_{\text{c}} = 100$ GeV are shown in \cref{tab: large-N GW peak amp}, alongside upper- and lower-bounds. A naive estimate of the peak amplitude scaling gives us  $h^2 \Omega_{\text{GW}}^{\text{peak}} \propto N^{-14/3}$. The only two quantities that scale with $N$ are the inverse duration $\tilde \beta \propto  N^{2} $ and the number of degrees of freedom $g_*\propto N^{2}$. In combination, the peak amplitudes scales as $h^2 \Omega_{\text{GW}}^{\text{peak}} \propto \tilde \beta^{-2} g_*^{-1/3} \propto N^{-14/3}$. This is in good agreement with our numerical fits given in \cref{tab: large-N GW peak amp}. The numerical scaling and the naive estimate agree quite well because of cancellations between the scaling of  $\tilde \beta$, which is not exactly $N^2$, and residual $N$ dependences in other GW parameters. It is worth highlighting that the peak amplitude scaling is mostly dominated by the inverse duration.

While we have a rather good understanding of the exponent of the $N$-scaling of the peak amplitude, the prefactor underlies a much larger uncertainty. This is due to the limit on supercooling discussed in \cref{sec:GW-spectrum F}, which we identified as the dominant uncertainty at large $N$. Concretely, at larger supercooling would lead to a smaller $\tilde \beta$ and a larger prefactor $A$ in the peak amplitude scaling \cref{Large-N GW peak amp func}.

In the right panel of \cref{fig: Large-N GW plots}, the peak frequency is plotted as a function of $N$ using both $T_{\text{c}} = 1$ GeV and $T_{\text{c}} = 100$ GeV. The explicit dependence of $f_{\text{peak}}$ on $T_{\text{c}}$ explains the approximate factor of $100$ difference between the two curves. The shape of the curves with $N$ is again dominated by $\tilde{\beta}$, and the extrema of the GW peak frequency curves remain approximately at $N \simeq 12$.

\begin{table}[t]
	\renewcommand{\arraystretch}{1.5}
	\begin{tabular}{|c|c|c|c|}
		\hline
		\multirow{2}{*}{$T_{\text{c}}$ }& $\tilde{\beta}$ & \multicolumn{2}{c|}{$h^2 \Omega_{\text{GW}}^{\text{peak}}$} \\
		\cline{2-4}
		& $B$ & $B$ & $\log_{10}(A)$ \\
		\hline
		1 GeV & $2.12^{+ 0.08}_{- 0.13}$ & $-4.56^{- 0.26}_{+ 0.34}$ & $-10.74^{- 0.36}_{- 0.20}$  \\[1ex]
		100 GeV & $2.15^{+ 0.07}_{- 0.11}$ & $-4.65^{- 0.23}_{+ 0.29}$ & $-10.57^{- 0.42}_{- 0.10}$ \\
		\hline
	\end{tabular}
	\caption{Fit parameters for the power-law in \cref{Large-N GW peak amp func} describing the large-$N$ ($N \geq 60$) behaviour of the inverse duration and the GW power spectrum peak amplitude. The superscripts (subscripts) of the parameter values were obtained by fitting using the upper- (lower-) bounds of $\tilde{\beta}$.}
	\label{tab: large-N GW peak amp}
\end{table}

In this work, we focused on the $SU(N)$ gauge group. Going beyond $SU(N)$, we expect the following two properties of the gauge group to influence how the GW spectrum changes: (I) the size of the group (the number of degrees of freedom, $n_\text{dof}$, of the adjoint representation), and (II) the centre symmetry group. The centre symmetry dictates the allowed terms in the effective Polyakov loop potential, see \cref{Potential general ansatz}. At small $N$, we believe these properties may both have a noticeable impact on the strength of the signal. The centre symmetry alters the shape of the effective potential and could enter non-trivially into the interface tension and latent heat, much like $n_{\text{dof}}$, which goes into fixing the model parameters. This will directly affect $\tilde{\beta}$, which dominates the signal. At large $N$, the system is dominated by $n_{\text{dof}}$, and for all classical Lie groups we have $n_{\text{dof}} \propto N^2$ in this regime. Considering also that a deconfined phase minimum temperature is a feature regardless of the group, meaning there is always a limit on supercooling, we expect similar large-$N$ scaling of the spectrum between gauge groups. However, new lattice studies are required to give a concrete answer to these questions.

\begin{figure}[t]
	\includegraphics[width=\linewidth]{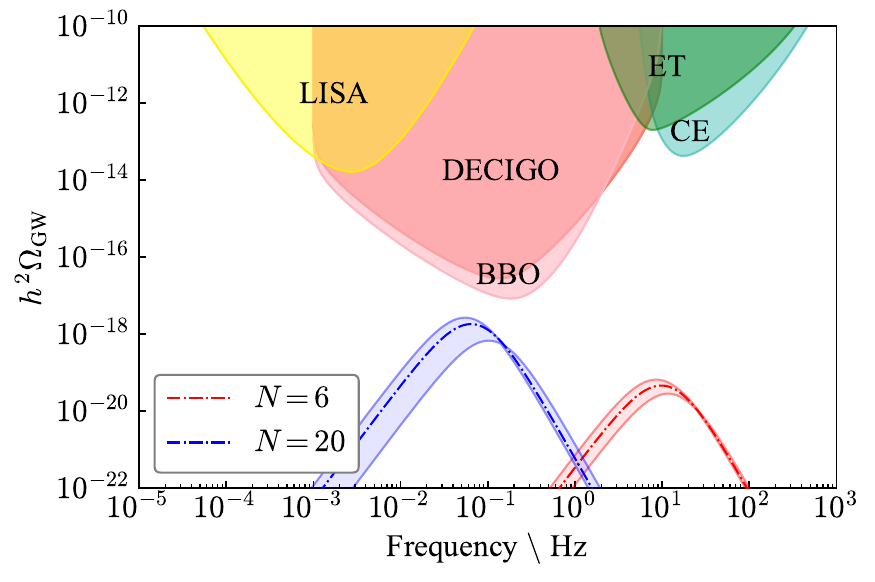}
	\caption{The stochastic GW power spectrum generated from the first-order confinement transition in $SU(N)$ pure Yang-Mills, shown for $N=6$ (red) and $N=20$ (blue). For $N=6$, a critical temperature of $T_\text{c}=100$ GeV was used, while $T_\text{c}=1$ GeV was used for the $N=20$ computation. The spectra are compared to the power-law-integrated sensitivity curves of the labelled GW detectors.}
	\label{fig: GW PS plot}
\end{figure}

\begin{figure*}[t]
	\includegraphics[width=.49\textwidth]{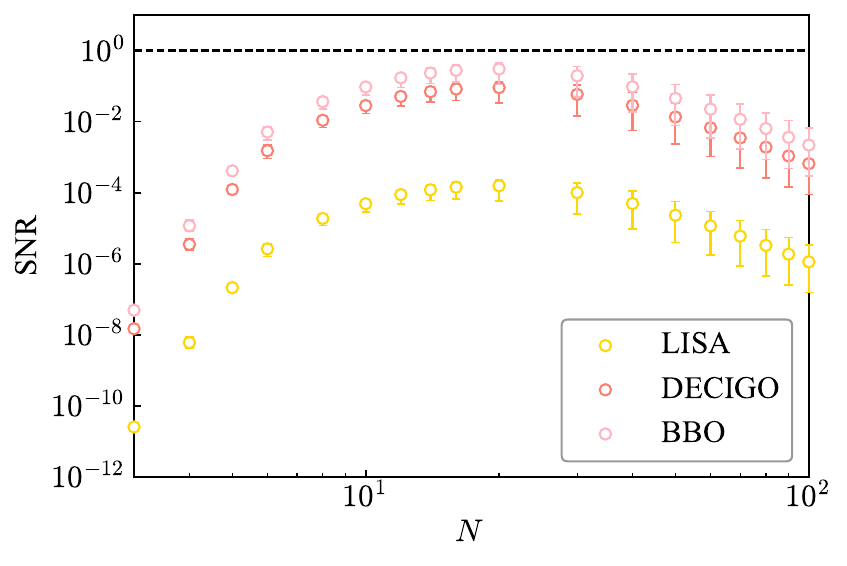}
	\hfill
	\includegraphics[width=.49\textwidth]{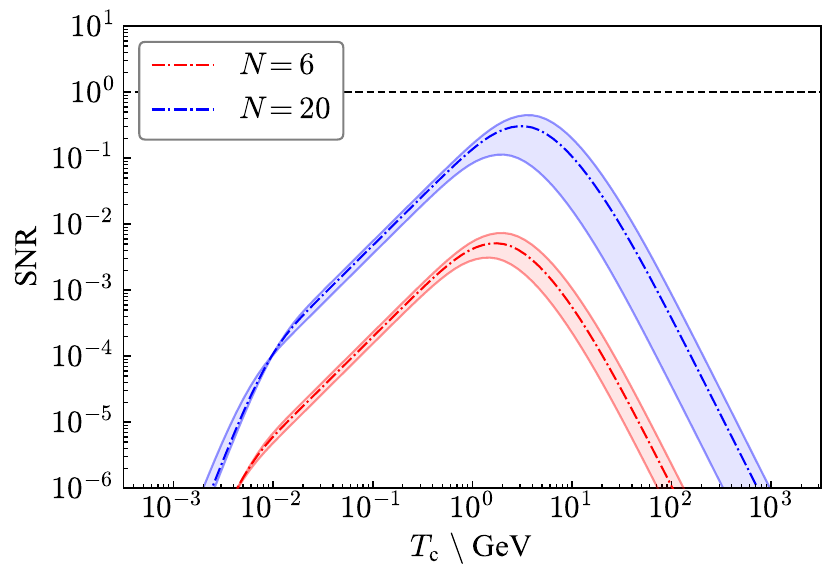}
	\caption{Signal-to-noise ratio for the confinement transition in an $SU(N)$ hidden sector. Left: as a function of $N$ evaluated at $T_{\text{c}} ^{\text{peak}}$; the critical temperature for which the peak of the $SU(N)$ GW power spectrum and the trough of the detector sensitivity curve align. Right: as a function of $T_\text{c}$ for $N=6$ and $N=20$ from the BBO experiment. The observation time is taken to be three years.}
	\label{fig:SNR}
\end{figure*}

\subsection{Detectability}
In \cref{fig: GW PS plot}, we plot the GW power spectrum as a function of frequency for $N = 6$ and $N = 20$ for $T_\text{c}=100$ GeV and $T_\text{c}=1$ GeV, respectively, alongside the power-law-integrated sensitivity curves for the LISA \cite{LISA:2017pwj, Baker:2019nia}, ET \cite{Punturo:2010zz, Hild:2010id, Sathyaprakash:2012jk, ET:2019dnz}, CE \cite{LIGOScientific:2016wof, Reitze:2019iox}, DECIGO \cite{Isoyama:2018rjb, Yagi:2011wg, Kawamura:2006up, Seto:2001qf}, and BBO \cite{Crowder:2005nr, Corbin:2005ny, Harry:2006fi} experiments, see \cite{Schmitz2021} for a nice summary of all sensitivity curves. A better test for the detectability of a signal is given by the SNR, which is detector-dependent, but plots of the form shown in \cref{fig: GW PS plot} are useful for understanding the wider detector landscape. We plot for $N = 6$ due to the applicability of the thin-wall approximation, and for $N = 20$ as it corresponds to the most detectable signal we found. At a critical temperature of $T_{\text{c}} = 100$ GeV, the GW power spectrum from the $SU(6)$ phase transition, which aligns within the bounds of ET and CE, is undetectable. That being said, the same would be true if the transition occurred at $T_{\text{c}} = 1$ GeV, as its peak amplitude is approximately three orders of magnitude lower than the peak of the BBO detector sensitivity. The peak amplitude of the power spectrum from the $SU(20)$ confinement phase transition is significantly greater, yet likewise remains undetectable. A critical temperature slightly above $T_{\text{c}} = 1$ GeV appears to be optimal for its chances of detection. In contrast with the results of \cite{Huang2021}, we have found a significant reduction in the amplitude of the GW power spectrum. This is due to both the refinements made to the model itself and the more accurate estimates of the bubble wall velocity and, in turn, the efficiency factor.

In \cref{fig:SNR} we present the SNR of the phase transition at the three most sensitive detectors we considered, namely LISA, DECIGO, and BBO, for various values of $N$ and $T_{\text{c}}$. The left-hand plot shows the SNR as a function of $N$ at each of these detectors, evaluated at $T_{\text{c}}^{\text{peak}}$. This is the value of the critical temperature for which the peak frequency of $h^2 \Omega^{\text{GW}}(f)$ aligns exactly with the trough of the detector sensitivity curve. Therefore, this is the optimal value of $T_{\text{c}}$, and the SNRs shown are of the best-case scenario for detection. Here, we make the optimistic assumption that a signal with SNR $>$ 1 is detectable. Despite this, we see that each signal predicted using the PLM is undetectable at the GW observatories. This is true even within the error bars, however, as is discussed in \cref{sec:Gw-spectrum-errors}, we believe the error bars to be an underestimate for larger values of $N$. Around the SNR peak of $N \simeq 20$, a broader error estimate could imply a possibly detectable signal.

In the right panel of \cref{fig:SNR}, we display the SNR at BBO for $N = 6$ and $N = 20$ as a function of $T_{\text{c}}$. At BBO, both signals are peaked around $T_{\text{c}}^{\text{peak}} = 1 - 10$ GeV. Moving away from the peak, if the critical temperature were to change by a factor of ten, we see suppression of the SNR by an order of magnitude in the best case. So, while it could be that the possible detectability of, for instance, the $N = 20$ phase transition is restored with a better error estimate, one would still require the critical temperature to be approximately equal to $T_{\text{c}}^{\text{peak}}$ for hope of a detection.

Our analysis suggests that planned and future GW observatories will not probe the parameter space of a first-order confinement phase transition in an $SU(N)$ dark sector. However, while an $SU(N)$ dark sector is a natural extension of the Standard Model, it does not provide a viable single-component dark matter candidate since the corresponding parameter space of dark glueballs is largely ruled out \cite{Asadi:2022vkc, Carenza:2022pjd, Carenza:2023eua}. In this light, our work provides conceptual development of a more rigorous treatment of a strongly-coupled dark phase transition, and paves the way towards studying more realistic QCD-like dark sectors, including fermions \cite{Reichert:2021cvs, Pasechnik2024, Houtz:2025ogg, Schwaller:2015tja}. We look forward to implementing improved lattice fits through modified kinetic terms, and a consistent wall velocity in these models.

\subsection{Discussion of Errors}
\label{sec:Gw-spectrum-errors}

In this section, we evaluate the accuracy of the error estimates given by the error bars in \cref{fig: Large-N GW plots,fig: GW PS plot,fig:SNR}. These upper- and lower-bounds were obtained from the fitting and matching procedures described in \cref{sec:PLM B}, and correspond to the uncertainties quoted on the interface tension and latent heat from the lattice. We believe these error estimates to be sufficient where thin-wall applies, but expect them to underestimate the uncertainty at large $N$. There are other sources of error which are not included in these error bars that need to be illuminated.

Firstly, we comment again on the uncertainty associated with the minimum temperature at large $N$, this time in the context of the plots in \cref{fig: Large-N GW plots,fig:SNR}, see \cref{sec:GW-spectrum F} for a discussion of the GW parameters in this context. To summarise: in this work, we access large $N$ by a rescaling of the $N = 6$ effective potential \cref{V scaling}. This assumes that $N = 6$ is already large, and implies that the minimum temperature is a constant for $N\geq 6$, $\tau_{\text{min}}^{N=6} = \tau_{\text{min}}^{N \geq 6}$. If this assumption holds, then the presented error bars at large $N$ are accurate. Conversely, if $\tau_{\text{min}}^N$ settles at a lower value than $\tau_{\text{min}}^{N=6}$, we would find a larger degree of supercooling and thus a larger GW signal at large $N$. In this respect, our results at large $N$ of the GW spectrum peak amplitude and SNR should be taken as a lower bound, but given we expect the true minimum temperature to be only slightly lower than $\tau_{\text{min}}^{N=6}$, the resulting signals would remain weak. In summary, this implies that we might have underestimated the upper error bar at large $N$. Independent of the degree of supercooling at large $N$, the results presented in this work act as a useful guide for the expected behaviour in the GW power spectrum at large $N$. 

The results at small $N$ are more accurate, given the agreement with the thin-wall approximation, and we believe the error bar to be a better estimate. However, there are hidden sources of uncertainty that the error band does not take into account. To estimate the wall velocity in this work we employed the novel method presented in \cite{SanchezGaritaonandia2024}, which applies in theories where there is a large jump in the number of degrees of freedom across the phase transition. In $SU(N)$ Yang-Mills theory, where this change scales as $N^2$, the large-jump assumption only strictly holds for large values of $N$. This introduces potentially significant uncertainty in the wall velocity at small $N$, which is not accounted for in the error bar. As $N$ increases, the large-jump assumption holds more strongly and this uncertainty drops. We also highlight that this directly impacts the uncertainty associated with the efficiency factor, which is inherently uncertain given the choice to use a constant sound speed and the fits given in \cref{Kappa fits}. We expect our estimates of $\kappa_{\text{sw}}$ are conservative, particularly for large $N$.

In the final computation of the GW power spectrum, there is one important phenomenon we have not taken into account, namely, the reheating of the plasma. The problem is nicely summarised in \cite{Athron:2023xlk}. The general idea is the following: in a supercooled phase transition, the difference in free energy between the false and true vacua is released. This energy can manifest itself in various forms, an example of which is the kinetic energy of the plasma, which is given by $\kappa_{\text{sw}}$, and indeed it can also go into reheating the plasma. In \cite{Ajmi:2022nmq}, a detailed study of the impact of this reheating effect in FOPTs proceeding through subsonic deflagrations was performed. It was found that the reheating of the plasma in front of the bubble wall decreases the rate of further bubble nucleation. Notably, the effect was enhanced for greater values of $\tilde{\beta}$ and $\alpha$. In the end, the suppressed bubble nucleation increased the average separation of nucleated bubbles, leading to more violent collisions. The overall effect was to increase the amplitude of the GW spectrum. We expect this to have a noticeable effect in the $SU(N)$ confinement phase transition studied here; the PLM generally exhibited very large values of $\tilde{\beta}$, which strongly suppressed the signal. We also point out the $N^2$-scaling of the latent heat and relatively small degree of supercooling. Furthermore, in \cite{Cutting:2019zws}, reheating of the plasma during bubble wall collisions is identified to cause the reformation of droplets of the metastable phase, which decreases the bubble wall velocity. This effect would act to suppress the final GW signal. One therefore requires a proper treatment of this phenomenon, which was beyond the scope of this work, and we postpone it to future studies.

\section{Conclusions}
\label{sec:conclusions}

We have presented a detailed computation of the gravitational wave signal generated during the confinement phase transition in $SU(N)$ pure Yang-Mills theory. Our work refines the effective Polyakov loop model developed in \cite{Huang2021} by incorporating recent lattice results, which clarify the large-$N$ scaling of the interface tension. Specifically, the identification of $\sigma \propto N^2$ scaling from the lattice \cite{Rindlisbacher:2025dqw} led to the introduction of a $\delta N^2$ prefactor for the kinetic term to reproduce this feature in the effective model, see \cref{fig: Sigma and Lh plots}. The parameter $\delta$ was tuned to precisely match these results, but remained of the same order for all $N$.

We compared the results of the phase transition inverse duration, $\tilde{\beta}$, from the Polyakov loop model to the thin-wall approximation. For small values of $N$, specifically $N \leq 6$, we found strong agreement, given the exact matching of the PLM to lattice data and similar predictions for the (small) degree of supercooling. At large $N$ we demonstrated the breakdown of the thin-wall approximation, which is attributed primarily to the presence of a minimum temperature of the deconfined phase predicted only in the effective model. The comparison between the PLM and the thin-wall approximation is shown in \cref{fig: Beta plots}. The agreement with thin-wall at small $N$ is a notable result of this work.

To estimate the wall velocity, we employed the novel framework \cite{SanchezGaritaonandia2024}, which assumes a large change in the number of degrees of freedom at the phase transition. Particularly in the large-$N$ limit, this approach is applicable to the $SU(N)$ Yang-Mills theory, which has $\Delta n_{\text{dof}} \propto N^2$. We found wall velocities comparable to those obtained in \cite{SanchezGaritaonandia2024} using the dark holographic model presented in \cite{Guersoy2009}. The identification of a suitable estimate for the wall velocity is a significant improvement made compared to previous studies \cite{Helmboldt:2019pan, Kang:2021epo, Reichert:2021cvs, Huang2021, Morgante2023, Halverson2021, Pasechnik2024, Ares:2020lbt}, which generally have left it as a free parameter.

We analysed the large-$N$ behaviour of the gravitational wave parameters and, in turn, that of the GW spectrum. In the large-$N$ limit, we found that the nucleation temperature approached the minimum temperature of the deconfined phase. This meant quantities with no explicit $N$-dependence, such as the strength parameter $\alpha$, and wall velocity, $\xi_{\text{w}}$, also became constant in this limit. The inverse duration exhibited $\tilde{\beta} \propto N^{2.12}$ scaling in this regime, in rough agreement with the naive $N^2$ estimate. This quantity was found to dominate the analysis.

A striking result is the presence of a turning point in the GW peak amplitude at $N \simeq 20$, related to the turning point of $\tilde{\beta}(N)$ at $N \simeq 12$, which is a consequence of competing effects. At small $N$, the thin-wall approximation holds, and the behaviour of the inverse duration follows from $\tilde{\beta} \propto L/\sigma^{3/2}$ \cite{Reichert:2021cvs}, such that approximately $\tilde{\beta} \propto 1/N$. At large $N$, the thin-wall approximation is not valid and $T_{\text{n}} \simeq T_{\text{min}}$ is being approached. The nucleation temperature is effectively constant, meaning $S_3/T$ and $\tilde\beta$ are dominated by $N^2$, which comes from the number of degrees of freedom. The turning point appears between these regimes and signals the onset of maximal supercooling.

For all $N$, we obtained rather weak gravitational wave signals, primarily due to the large suppression from $\tilde{\beta}$. The strongest signal we found was that from the $N=20$ phase transition, which itself is undetectable at future experiments. However, due to certain assumptions made in the computations, we expect these to be conservative results. A notable result of this work was to demonstrate the suppression of the GW signal at large $N$.

To estimate the uncertainty on the predictions for the GW spectrum, we incorporated the error bars on the lattice results for the interface tension and latent heat in our calculation. While we expect this to be a reasonable error estimate at small $N$, this is an underestimate of the uncertainty for larger values of $N$. A number of sources of uncertainty were not quantified in the error bar, such as the reheating of the plasma, which was not considered in this work. A significant source of error at large $N$, which was discussed extensively, is the minimum temperature of the deconfined phase. New lattice studies are required to investigate this further.

Overall, we have provided conceptual development to accurately determine the GW spectrum of an $SU(N)$ confinement phase transition. This was achieved by improving the effective Polyakov loop model such that it accurately reproduces all lattice data, including novel surface tension results at large $N$. We have determined that the GW signal will not be detectable at future GW observatories, and the peak amplitude rapidly decays at large $N$, $h^2 \Omega_{\text{GW}}^{\text{peak}} \propto N^{-14/3}$. The weak signals are a consequence of the inherent strong coupling, which leads to a rapidly changing effective potential. This in turn destabilises the metastable vacuum close to the critical temperature, leading to a small maximum degree of supercooling, $1 - T_\text{min}/T_\text{c} \ll 1$. Despite this small supercooling, we demonstrate that the thin-wall approximation breaks down at large $N$. This is as a result of the disappearance of the barrier at $T_{\text{min}}$. At temperatures slightly above this, where the barrier is small, the nucleated bubbles are small, and the thin-wall approximation is not applicable.

Our work paves the way for studies of dark sectors described by different symmetry groups, and more realistic QCD-like sectors containing fermions in various representations. Further lattice results for quantities such as the interface tension, latent heat, or direct information on the potential would be welcome to constrain the corresponding effective models. In particular, QCD-like sectors close to the conformal Banks-Zaks window might offer intriguing features where the GW signal is enhanced.

\bigskip

\centerline{\bf Acknowledgements}
We thank Mark Hindmarsh, Biagio Lucini, David Mason, Maurizio Piai, Davide Vadacchino, Jorinde van de Vis, Zhi-Wei Wang, and Fabian Zierler for discussions. This work is supported  by the Science and Technology Research Council (STFC) under the Consolidated Grant ST/X000796/1, the Ernest Rutherford Fellowship ST/Z510282/1, and the STFC Studentship Grant ST/Y509620/1.

\bibliographystyle{mystyle}
\bibliography{GW-SUNrefs}

\end{document}